\documentclass[12pt]{article}

\usepackage{graphicx}
\usepackage{multicol,multirow}
\usepackage{amsmath,amssymb,amsfonts}
\usepackage{mathrsfs}
\usepackage{amsthm}
\usepackage{booktabs}
\usepackage{rotating}
\usepackage{appendix}
\usepackage{ifpdf}
\usepackage[T1]{fontenc}
\usepackage{textcomp}
\usepackage{xcolor}
\usepackage{lipsum}
\usepackage[colorlinks,allcolors=blue]{hyperref}
\usepackage{makecell}
\usepackage{CJKutf8}
\usepackage{lscape}
\usepackage{graphicx,psfrag,epsf}
\usepackage{enumerate}
\usepackage{color}
\usepackage{natbib}
\usepackage{url} 

\newcommand{\blind}{0}

\newcommand{\bd}[1]{\mbox{\boldmath$#1$}}

\begin{document}

\def\spacingset#1{\renewcommand{\baselinestretch}%
{#1}\small\normalsize} \spacingset{1}

\thispagestyle{empty}
\if1\blind
{
  \maketitle
} \fi

\if0\blind
{
  \bigskip
  \bigskip
{\bf{}}
  \bigskip
  \begin{center}
    {\Large\bf Matrix Decomposition-Based Approach to Estimate the STARTS Model}
\end{center}
} \fi
\bigskip
\begin{center}
{\large{{\bf{Satoshi Usami}}}}\\
{\small{Graduate School of Education, The University of Tokyo, 7-3-1, Hongo, Bunkyo-ku, Tokyo, Japan.}}\par
\bigskip
usamis@g.ecc.u-tokyo.ac.jp 
\end{center}
\bigskip
\bigskip
\bigskip \hspace{5mm}
Competing Interests: The author declares none.\par
\bigskip
\bigskip
Financial Support: This research was supported by Japan Society for the Promotion of Science\par KAKENHI
Grant Number JP23K02861.\par
\bigskip
\bigskip
Data Availability: Data used in empirical example are available from the author upon request.
\clearpage
\setcounter{page}{1}
\newpage
\if1\blind
{
  \maketitle
} \fi

\if0\blind
{
  \bigskip
  \bigskip
{\bf{}}
  \bigskip
  \begin{center}
    {\Large\bf Matrix Decomposition-Based Approach to Estimate the STARTS Model}
\end{center}
} \fi
\bigskip
\begin{center}

{\small{}}
\end{center}
\bigskip
\begin{abstract}
We propose a new estimation method for the Stable Trait, Auto Regressive Trait, and State (STARTS) model, which is well known for its frequent occurrence of improper solutions. The proposed approach is implemented through a two-stage estimation procedure that combines matrix decomposition factor analysis (MDFA) based on eigenvalue decomposition with conventional SEM estimation principles. By reformulating the STARTS model within a factor-analytic framework, this study presents a novel way of applying MDFA in the context of structural equation modeling (SEM). Through a simulation study and an empirical application to ToKyo Teen Cohort data, the proposed method was shown to entail a substantially lower risk of improper solutions than commonly used maximum likelihood, conditional ML, and (unweighted) least squares estimators, while tending to yield solutions similar to those obtained by ML. Compared with Bayesian estimation, the proposed method does not require the specification of appropriate (weakly informative) prior distributions and may effectively mitigate bias issues that arise when the number of time points is small. Applying the proposed method, as well as conducting sensitivity analyses informed by it, will enable researchers to more effectively delineate the range of plausible conclusions from data in estimating the STARTS model and other SEMs.
\end{abstract} 
\noindent%
\vfill

\newpage
\spacingset{1.6} 
\section{Introduction}
\label{sec:intro}
It is estimated that more than 30,000 longitudinal studies are published annually worldwide, and this number is expected to continue increasing (Usami, 2026). Among this massive volume of research, a substantial proportion aims to examine longitudinal relations among variables—that is, the dynamics of change over time. Particularly in psychology, growing attention has been directed toward modeling within-person relations while simultaneously accounting for the influence of stable individual differences by modeling stable traits (or, unit effects) as latent variables. For example, in personality research, analysis models often assume the existence of stable traits to evaluate the long-term stability and short-term variability of personality characteristics such as extraversion and neuroticism using longitudinal data. The widespread adoption of the random intercept cross-lagged panel model (RI-CLPM; Hamaker, Kuiper, \& Grasman, 2015), which aims to examine reciprocal within-person relations among variables, is emblematic of this trend.

On the other hand, psychology has long employed models that explicitly incorporate measurement error in addition to stable traits, such as the stable trait–autoregressive trait–state (STARTS) model (Kenny \& Zautra, 1995, 2001). 
This model, which is more general and is expected to be more closely aligned with the realities of psychological measurement, was proposed earlier than the RI-CLPM and continues to be widely used across various domains of \textcolor{black}{psychological research, including educational, personality, clinical, and developmental psychology}. \textcolor{black}{Whereas the RI-CLPM is typically formulated as a bivariate model involving two variables ($x$ and $y$), with primary interest in examining their reciprocal relations, the STARTS model is often formulated as a univariate model focusing on a single variable ($y$).} The STARTS model decomposes individual differences in measurements across time into three sources of variation: a time-invariant stable component (i.e., stable traits factor), a time-varying autoregressive component (i.e., within-person fluctuations), and an occasion-specific state component (i.e., measurement error). For example, Dicke et al. (2022) examined changes in Australian school principals’ emotional exhaustion (i.e., burnout) using longitudinal data ($N$ = 5,509; $T$= 8) analyzed with the STARTS model. Their results demonstrated that the variance in principals’ emotional exhaustion was distributed almost evenly between a persistent autoregressive component (reflecting within-person fluctuations) and a stable trait component (reflecting stable traits factor), with a somewhat smaller portion attributable to occasion-specific fluctuations (reflecting measurement error). Furthermore, variability in these structural profiles was found to be primarily linked to individual attributes of the principals—such as years of experience and gender—rather than work-related contextual factors (e.g., school type or educational stage).

In light of the recent proliferation of intensive longitudinal data (ILD; e.g., McNeish \& Hamaker, 2020), further applications of the STARTS model—as well as its extended variants—are expected in the near future. \textcolor{black}{Relatedly, Nestler and Humberg (2024) showed that the STARTS model can be formulated as a mixed-effects model and estimated using the \texttt{nlme} package in R. Nestler et al (2025) embeded the single- and multiple-indicator STARTS model within the dynamic SEM framework.}

Models such as the RI-CLPM and the STARTS model are sometimes referred to collectively as residual-level models (e.g., Andersen, 2022), as they describe dynamic processes operating on latent deviations (rather than observed scores). All of these can be formulated within the structural equation modeling (SEM) framework, and maximum likelihood (ML) estimation is most commonly used especially for the RI-CLPM. However, there are cases in which models other than residual models provide a better fit to empirical data (e.g., Usami, 2022), and \textcolor{black}{when two or more variables are considered,} modeling latent interactions in residual models often introduces additional implementation challenges (e.g., Usami, 2023). Consequently, the selection of an appropriate longitudinal model remains an active area of methodological discussion (e.g., Hamaker, 2023; Lucas, 2023; L\"{u}dtke \& Robitzsch, 2022, 2025; Orth et al., 2021; Usami, 2021; Usami, Murayama, \& Hamaker, 2019). Although it should be kept in mind that such more fundamental theoretical discussions are still ongoing, it is nevertheless important to note that the STARTS model, in particular, capitalizes on the advantages of residual models by separating residuals in the autoregressive component from measurement error, thereby allowing measurement error to be represented explicitly. 

Although incorporating measurement error is conceptually natural for most psychological measurement contexts, 
the major obstacle to broader adoption of the STARTS models lies in the frequent occurrence of improper solutions (e.g., negative error variances, non-positive definite covariance matrices among trait factors), non-convergence, and unstable parameter estimates. For example, Usami, Todo, and Murayama (2019) reported that while improper solutions can also occur in the RI-CLPM, they are far more prevalent in the STARTS models—even when the true data-generating process corresponds exactly to the STARTS structure. 

Building on Lüdtke et al. (2018), the causes of these frequent estimation problems can be summarized as four interrelated aspects:
(i) Highly dependent parameter estimates and a nearly singular Hessian matrix, (ii) Sensitivity to small changes in the sample covariance matrix, (iii) Parameter regions close to non-identification, and
(iv) Empirical under-identification and flat likelihoods. First, because the three kinds of variance components (stable trait, autoregressive trait, and state variance) and the autoregressive coefficient are highly interdependent, ML estimation often encounters a nearly singular Hessian matrix. Second, this strong interdependence leads to large standard errors and instability of estimates—very small sampling fluctuations in the covariance matrix can induce large changes in parameter estimates, making ML estimation numerically unstable (Kenny \& Zautra, 2001). Third, the STARTS model becomes particularly unstable when population parameters approach boundary conditions, specifically when the autoregressive variance is small or when the standardized stable-trait variance is near 0 or 1. Fourth, even when the model is technically identified, the likelihood surface can be flat with respect to variance parameters, making it difficult for optimization algorithms to find a unique solution—a phenomenon empirically demonstrated by Lüdtke et al. (2018, Figure 5), who showed that a broad range of parameter values can yield nearly identical likelihoods.

These issues are mutually interrelated, and depending on conditions such as the number of time points and population parameter values, avoiding improper solutions can be extremely difficult. One possible countermeasure is constrained estimation like constrained ML (CML), in which the optimization space is restricted to admissible values (e.g., variances constrained to be positive). However, constrained estimation frequently produces boundary values (i.e., estimates at the lower or upper bound), leading to biased estimates (Searle, Casella, \& McCulloch, 1992; cited from Lüdtke et al., 2018). Moreover, \textcolor{black}{recent simulation results for confirmatory factor analysis (CFA) suggest that several non-iterative estimators based on least squares (LS) can avoid Heywood cases (Dhaene \& Rosseel, 2023).}

When improper solutions occur in the estimation of the STARTS model, it is often the case that researchers alter model specifications (e.g., by imposing equality constraints on parameters) or modify the data being analyzed (e.g., by excluding specific waves or participants) in order to obtain admissible solutions. As an alternative to such partly conservative approaches, Bayesian estimation is also useful and has been employed as a procedure for addressing improper solutions. Lüdtke et al. (2018), demonstrated that when appropriate (weakly informative) priors are specified, a Bayesian approach can effectively prevent improper solutions and stabilize parameter estimation in the STARTS model. Bayesian estimation is also well known as an effective strategy for avoiding improper solutions in hierarchical linear models, particularly those including random coefficients. \textcolor{black}{Nestler et al. (2025) also used Bayesian estimation in Mplus when embedding the STARTS model within the dynamic SEM framework}.

However, Bayesian estimation is known to carry the risk of inducing parameter bias; for example, in the context of mediation models, it has been reported that under small sample sizes, Bayesian estimation tends to yield more stable results than CML, but often at the cost of increased bias (e.g., Smid et al., 2020; Ulitzsch et al., 2023). Moreover, under such small-sample conditions, estimation results can be highly sensitive to the choice of prior distributions (e.g., Miočević et al., 2021). Therefore, sensitivity analyses comparing results under different prior specifications are commonly recommended. Yet, because the Bayesian approach for the STARTS model typically relies on Markov chain Monte Carlo (MCMC) algorithms to approximate posterior distributions, computational costs are usually higher than for ML, and these costs increase further when sensitivity analyses are performed. Furthermore, even when plausible prior distributions are specified, particularly when the prior information is weak, the resulting estimates may still correspond to the zero boundary, yielding solutions similar to those obtained by CML.

In the present study, we propose a new estimation method to address the problem of improper solutions in the STARTS model. The proposed approach is implemented through a two-stage estimation procedure that combines matrix decomposition factor analysis (MDFA; Adachi \& Trendafilov, 2018) based on eigenvalue decomposition with conventional SEM estimation principles. MDFA is an approach based on matrix algebra, and estimate parameters by iterative estimation using a LS criterion based on data space rather than covariance space as SEM. MDFA was originally proposed in the context of factor analysis, and recent developments include its extensions to SEMs involving structural relations among latent variables (Yamashita, 2024) and theoretical examinations of its estimator properties (Terada, 2025). One of the key features in MDFA is that negative variance estimates do not occur in principle during iterative updates. In the present study, by reformulating the STARTS model within a factor-analytic framework, we presents a novel way of applying MDFA in the context of SEM. 

To date, the application of MDFA to longitudinal models—including the STARTS model—has not been reported, and its estimation performance and effectiveness in addressing the problem of improper solutions remain unexplored. Through a simulation study and an empirical application to ToKyo Teen Cohort (TTC) data, we will show that the proposed method entails a substantially lower risk of improper solutions than ML, CML and ULS estimators, and that it may effectively mitigate bias issues that can arise especially in Bayesian estimation when the number of time points is small. 

The structure of this paper is as follows. Section 2 provides an overview of the STARTS model and describes ML, CML, and ULS estimators. Section 3 introduces MDFA and shows how it can be applied to estimate parameters in the STARTS model. Section 4 presents a simulation study for evaluating the frequency of improper solutions, bias and RMSE of estimates, and computational cost across different estimation methods. Section 5 applies the proposed method to the TTC data. Section 6 concludes with a summary and directions for future research.
\section{STARTS Model and Estimation}
\subsection{STARTS model}
\begin{figure}[tbp]
  \centering
  \includegraphics[width=\linewidth]{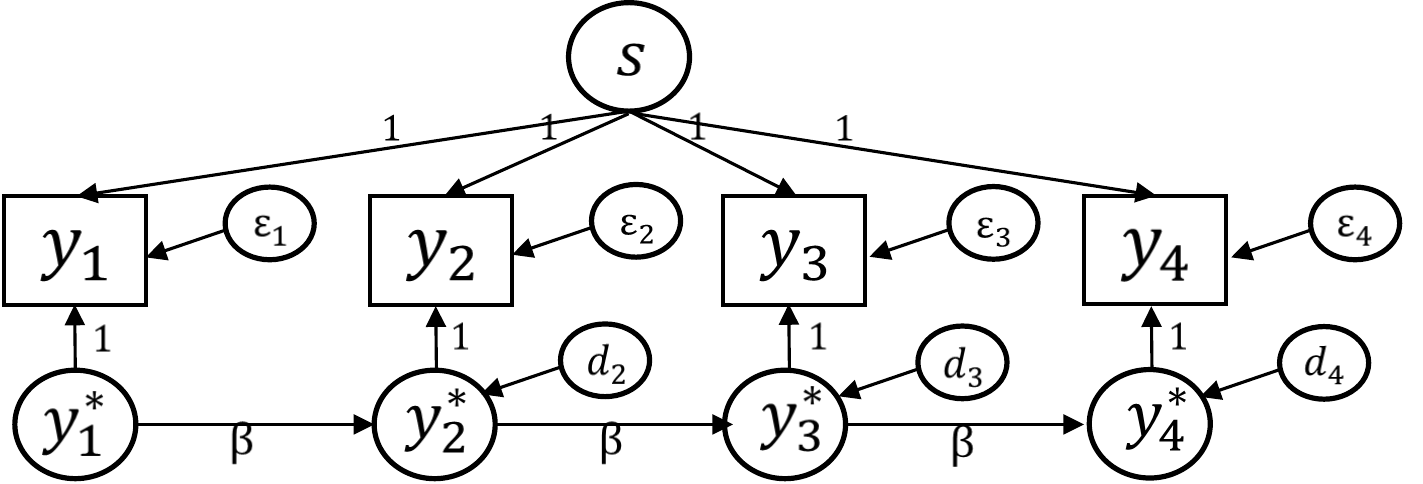}
  \caption{Path diagram of the STARTS model ($T=4$)}
  \label{fig:pathdiagram}
\end{figure}

Suppose that data are generated at fixed and equally spaced time points, and let $y_{it}$ denote a continuous observed variable at time point $t$ ($t=1,\dots,T$) for person $i$ ($i=1,\dots,N$). For expository purposes, here we consider the case in which there is a single focal variable ($y$).

In the STARTS model (Figure~1 for the case of $T=4$), $y_{it}$ is considered to consist of latent true scores ($f_{it}$) and measurement errors ($\epsilon_{it}$), that is,
\begin{gather}
y_{it}=f_{it}+\epsilon_{it}, \label{meaeq}
\end{gather}
for all time points. These measurement errors are usually assumed to be normally distributed (with zero mean and variance of $\psi^2$) and mutually uncorrelated.

The latent true scores $f_{it}$ are then modeled as
\begin{gather}
f_{it}=\mu_{t}+s_{i}+y^*_{it} \label{deceq}
\end{gather}
for all time points. This equation is used to decompose latent true scores $f_{it}$ into a temporal group mean at $t$ ($\mu_{t}$),
a stable trait factor of person $i$ ($s_{i}$), and a temporal deviation ($y^*_{it}$). Since $\mu_{t}$ is not of substantive interest here (i.e., mean structure is irrelevant to the current discussion), and in the subsequent section for the proposed method, we assume $y_{it}$ is centered by group mean (i.e., $\mu_{t}=0$). Also, we assume both the trait factor and temporal deviation have zero mean, and they are uncorrelated with measurement errors. \textcolor{black}{The variances of $s_{i}$ and $y^*_{it}$ are expressed here by 
$\phi^2$ and $\sigma_w^2$, respectively. In this way, the variance of the temporal deviation ($y^*_{it}$) is typically assumed to be constant across time points, as are $\psi^2$ and $\phi^2$.}

The stable trait factor ($s_i$) exhibits the between-person or stable individual differences, and the temporal deviation ($y^*_{it}$) represents the within-person variability from an expected value for person $i$ at $t$ (i.e., $\mu_{t}+s_{i}$). The initial within-person variability ($y^*_{i1}$) is modeled as an exogeneous variable.

Importantly, the stable trait factor is assumed to be uncorrelated with the within-person variability:
\begin{gather}
Cov(s_{i},y^*_{it})=0
\end{gather}
for each $i$ and $t$. Consequently, as with the RI-CLPM, the stable trait factor $s$ has only direct effects on outcomes, and
each latent true score can be decomposed into a (linear) sum of time-invariant ($s_{i}$) and time-varying ($y^*_{it}$) factors that are mutually uncorrelated.

Within-person variability $y^*_{it}$ is further modeled as
\begin{flalign}
y^*_{it}&=\beta y^*_{i(t-1)}+d_{it} \label{reg}
\end{flalign}
for $t\geq 2$. In Equation~(\ref{reg}), the (first-order) autoregressive process is assumed to represent the within-person relation. $\beta$ \textcolor{black}{($0<|\beta|<1$)} is an (time-invariant) autoregressive
coefficient, and $d_{it}$ denotes a residual that is uncorrelated with $y^*_{it'}$ ($t'\leq t-1$) and $s_i$. \textcolor{black}{Here, the time-invariant residual variance is assumed to satisfy
$Var(d_{it})=\omega^2$. Furthermore, given
$Var(y^*_{it})=\sigma_w^2$, the constraint
\begin{gather}
\omega^2=(1-\beta^2)\sigma_w^2\label{within}
\end{gather}
is imposed. Thus, $\omega^2$ is expressed as a function of $\beta$ and $\sigma_w^2$.}\footnote{In the \textcolor{black}{bivariate} STARTS model, \textcolor{black}{in the same way,} it is common to impose appropriate nonlinear constraints on the parameters so that the variances of the trait factor, within-person variation (corresponding to the ‘autoregressive trait’ in the STARTS context), and measurement error (corresponding to the ‘state’ in the same context) remain time-invariant. In contrast, such constraints are typically not imposed in the RI-CLPM.}

From Equation (\ref{reg}), $y^*_{it}$ can be expanded as
\begin{align} 
y^*_{it}&=\beta y^*_{i(t-1)}+d_{it}\notag \\
&=\beta (\beta y^*_{i(t-2)}+d_{i(t-1)})+d_{it} \notag \\
&=\beta (\beta (\beta y^*_{i(t-3)}+d_{i(t-2)})+d_{i(t-1)})+d_{it}\notag\\
& \hspace{30mm}\vdots \notag \\
&\textcolor{black}{=\beta^{t-t'}y^*_{it'}+d_{it}+\beta d_{i(t-1)}+\beta^2d_{i(t-2)}+\dots+\beta^{t-t'-1}d_{i(t'+1)}},\label{withinexpand}\\
& \hspace{30mm}\vdots \notag \\
&=\beta^{t-1}y^*_{i1}+d_{it}+\beta d_{i(t-1)}+\beta^2d_{i(t-2)}+\dots+\beta^{t-2}d_{i2}. \label{withinexpand2}
\end{align}
\textcolor{black}{From the equation (\ref{withinexpand})}, the variance of $y^*_{it}$ and covariance between $y^*_{it}$ and $y^*_{it'}$ ($t'< t$) can be expressed as
\begin{align} 
\textcolor{black}{Cov(y^*_{it},y^*_{it'})}&\textcolor{black}{=\beta^{t-t'}Var(y^*_{it'})=\beta^{t-t'}\sigma_{w}^2}\label{wcov}
\end{align}
for $t\geq 2$.

Suppose $y_{it}$ is now mean-centered at each time point (i.e., $\mu_{t}=0$), from Equations (\ref{meaeq}) and (\ref{deceq}), $y_{it}$ can now be expressed as
\begin{align}
y_{it}&=s_{i}+y^*_{it}+\epsilon_{it} \label{starts1}.
\end{align}

The variance of $y_{it}$ and covariance between $y_{it}$ and $y_{it'}$ ($t'< t$) can be orthogonally decomposed as
\begin{gather}
Var(y_{it})=Var(s_i)+Var(y^*_{it})+Var(\epsilon_{it})=\textcolor{black}{\phi^2+\sigma_w^2+\psi^2,}\label{varst}\\
Cov(y_{it},y_{it'})=Var(s_i)+Cov(y^*_{it},y^*_{it'})+Cov(\epsilon_{it},\epsilon_{it'})=\textcolor{black}{\phi^2+\beta^{t-t'}\sigma_{w}^2.}\label{covst}
\end{gather}
The parameters in the STARTS model can be identified from data with $T \geq 4$. 
\subsection{Several estimators in SEM and improper solutions}
\subsubsection{Maximum likelihood estimation}
Suppose that outcomes are now mean-centered at each time point, and $\bd{y}_i=(y_{i1},\dots,y_{iT})^\top$ follow a multivariate normal distribution:
$\bd{y}_i\sim i.i.d.MVN(\bd{0},\bd{\Sigma}(\bd{\theta}))$, where 
$\bd{\Sigma}(\bd{\theta})$ is a $T \times T$ covariance structure 
expressed as in Equations (\ref{varst})-(\ref{covst}), which is
a function of unknown parameters $\bd{\theta}=(\psi^2,\phi^2,\beta,\textcolor{black}{\sigma_w^2})^\top$ \textcolor{black}{(as shown in Equation~(\ref{within}), $\omega^2$ is determined once $\sigma_w^2$ and $\beta$ are given).} 

The observed log-likelihood is given by $LL(\bd{\theta}|\bd{Y})=\sum_{i=1}^NLL_i(\bd{\theta}|\bd{y}_i)$, where, 
\begin{gather}
LL_i(\bd{\theta}|\bd{y}_i)=-\frac{T}{2}\log(2\pi)-\frac{1}{2}\log|\bd{\Sigma}(\bd{\theta})|-\frac{1}{2}\bd{y}^\top_i\bd{\Sigma}(\bd{\theta})^{-1}\bd{y}_i,
\end{gather}
and $\bd{Y}=(\bd{y}^\top_1,\dots,\bd{y}^\top_N)^\top$.
Instead of maximizing $LL(\bd{\theta}|\bd{Y})$, it is common to estimate 
$\bd{\hat{\theta}}$ by minimizing the following loss function based on the likelihood ratio test statistic
that is used for evaluating analysis models:
\begin{gather}
f_{ML}(\bd{\theta})=\mathrm{tr}(\bd{S}\bd{\Sigma}(\bd{\theta})^{-1})-\log|\bd{S}\bd{\Sigma}(\bd{\theta})^{-1}|-T,
\end{gather}
where $\mathrm{tr}$ denotes the trace of a matrix and $\bd{S}$ is a $T \times T$ (unbiased) sample covariance matrix of outcomes.
The ML estimator has been used most widely in SEM applications.
Hayakawa \& Sun (2022) showed that the ML estimator is
consistent and more efficient than other estimators (e.g., generalized LS) when both $N$ and $P$ (the number of variables, which is equal to $T$ in the current setting)
tend to infinity with $P/N \rightarrow c$ ($c$ is a finite constant with $0<c<1$).

It is common to perform iterative calculations using numerical methods like Newton–Raphson method to obtain parameter estimates $\bd{\hat{\theta}}$. Previous simulation studies and empirical applications of the STARTS model with ML estimation have indicated that estimation difficulties, including issues such as nonconvergence and improper solutions, often arise (L$\ddot{\mathrm{u}}$dtke et al., 2018). Another problem is that parameter estimates are often very unstable, which is reflected by large estimates of standard errors.

As noted above, one possible countermeasure is to use constrained estimation like CML. It is usually implemented by minimizing the $f_{ML}(\bd{\theta})$ subject to equality and boundary constraints on model parameters. Equality constraints are handled via reparameterization, whereas inequality constraints (e.g., non-negativity of variance parameters) are enforced through boundary-constrained numerical optimization. However, it is empirically known that boundary values can occur in CML and it can lead to biased parameter estimates (e.g., Searle et al., 1992).

\subsubsection{Bayesian estimation}
L$\ddot{\mathrm{u}}$dtke et al. (2018) introduced a general approach for stabilizing parameter estimation in the STARTS model through a Bayesian framework, making use of MCMC methods with (weakly informative) priors. 

Bayesian estimation treats all model parameters as random variables and evaluates their posterior distributions given the observed data. The posterior
distribution can be evaluated by
\begin{gather}
p(\bd{\theta}|\bd{Y})
\propto L(\bd{\theta}|\bd{Y})\,p(\bd{\theta}),
\end{gather}
where $L(\bd{\theta}|\bd{Y})$ represents the
likelihood and $p(\bd{\theta})$ denotes the
prior distribution.
 
L{\"u}dtke et al.\ (2018) demonstrated through simulations that Bayesian estimation with weakly informative prior distributions could avoid estimation problems while also exhibiting desirable properties in terms of bias and RMSE. However, \textcolor{black}{they also showed} that when the prior distribution is not specified appropriately (i.e., strongly deviates from the true parameter), the Bayesian approach can result in largely biased parameter estimates.

Ideally, a specification of informative priors should be accompanied by a sensitivity analysis that tests how sensitive the resulting parameter estimates are to different specifications of the priors (L$\ddot{\mathrm{u}}$dtke et al., 2018). Especially when the sample size is small, the prior specification has a greater influence on the estimation results, and researchers should seriously consider conducting a sensitivity analysis (L$\ddot{\mathrm{u}}$dtke, et al., 2018). However, conducting sensitivity analyses that involve changing various parameter settings can be complex and may lead to an increase in computational cost. They also noted that one limitation of the Bayesian approach was that it showed a tendency to produce coverage rates that were too high.

\subsubsection{Least squares estimation}
Another well-known estimation method in SEM is a weighted LS (WLS). Let $\bd{s}=vech(\bd{S})$ and $\bd{\sigma}(\bd{\theta})=vech(\bd{\Sigma}(\bd{\theta}))$ denote $T(T+1)/2 \times 1$ vectors that extract the unique (co)variance elements from $\bd{S}$ and $\bd{\Sigma}(\bd{\theta})$, respectively. When mean structure is not considered, the loss function of WLS can be written as:
\begin{gather}
f_{LS}(\bd{\theta})=(\bd{s}-\bd{\sigma}(\bd{\theta}))^\top \bd{\hat{W}}\,(\bd{s}-\bd{\sigma}(\bd{\theta})), \label{ULS}
\end{gather}
where $\bd{\hat{W}}$ is a weight matrix.

When the weight is removed (i.e., $\bd{\hat{W}}=\bd{I}$), this version of estimator is referred to as ULS. There are other types of LS estimators (e.g., Du \& Bentler, 2022; Hayakawa \& Sun, 2022; Zheng \& Bentler, 2021). Although ULS is known to be less efficient than other estimators and is neither scale-free nor scale-invariant (e.g., Kaplan, 2009), it is computationally inexpensive, straightforward to implement, and still yields consistent estimates. In addition, ULS does not require the extremely large sample sizes typically needed for WLS to obtain good quality point estimates, which makes it particularly useful in applications such as ordinal categorical data analysis (e.g., Rhemtulla et al., 2012) and stepwise estimation procedures in which the measurement model part is estimated first (e.g., Rosseel \& Loh, 2024).

In research contexts where the STARTS model is applied, sample sizes are not always large, and to the best of the present author’s knowledge, no study has systematically compared LS estimator with ML estimator in terms of estimation performance and the frequency of improper solutions in the STARTS model. For these reasons, the present study includes ULS in the comparison as part of an exploratory investigation. Furthermore, the principle of ULS estimation is also incorporated into certain part of the proposed method explained later.
\section{Proposed MDFA-Based Approach}
In this section, we first provide an overview of the MDFA framework. We then describe how this estimation approach can be utilized to the STARTS model, thereby introducing the proposed methodology.
\subsection{An overview of MDFA}
As its name suggests, MDFA has been discussed primarily within the context of factor analysis models. Following Adachi et al. (2019) and Yamashita (2024), we introduce the MDFA framework.

Let $\bd{Y}$ ($N\times T$) be a data matrix of $N$ observations with respect to $T$ observed variables (and is now column-centered). When MDFA is applied to $\bd{Y}$, the method is formulated by minimizing the following (Frobenius norm-based) LS criterion:
\begin{gather}
||\bd{Y}-(\bd{F}\bd{\Lambda}^\top+\bd{UD})||^2, \label{lossMDFA}
\end{gather}
with respect to the common factor score matrix $\bd{F} \,(N \times r)$, 
the unique factor score matrix $\bd{U} \,(N \times T)$, the factor loading matrix $\bd{\Lambda} \,(T \times r)$, and the diagonal matrix $\bd{D} \,(T \times T)$ whose diagonal elements contain the square roots of the unique variances. Here $r$ indicates the number of assumed common factors. \textcolor{black}{$||\cdot||$ denotes the Frobenius norm, defined as the square root of the sum of the squared elements of a matrix; thus, the criterion in (\ref{lossMDFA}) measures the overall squared discrepancy between the observed data matrix and its model-based approximation.}

We now assume that the variances of the factor scores in $\bd{F}$ and $\bd{U}$ are equal to one, and that these factors are mutually uncorrelated (i.e., $\frac{1}{N}\bd{F}^{\top}\bd{F} = \bd{I}_r, \frac{1}{N}\bd{U}^{\top}\bd{U} = \bd{I}_T, \bd{F}^{\top}\bd{U} = \mathbf{0}_{r \times T})$. Furthermore, both $\bd{F}$ and $\bd{U}$ are column centered.

\textcolor{black}{Let $\bd{Z} = (\bd{F}, \bd{U})$ denote the matrix collecting the factor scores, and let $\bd{B} = (\bd{\Lambda}^{\top}, \bd{D})$ denote the matrix associated with the model parameters (except for the factor scores).} Then, the above criterion can be rewritten as
\begin{gather}
||\bd{Y}-(\bd{F}\bd{\Lambda}^\top+\bd{UD})||^2 =||\bd{Y}-\bd{Z}\bd{B}^\top||^2 =||\bd{Y}-\bd{Z}\bd{S_{YZ}}^\top||^2+N||\bd{S_{YZ}}-\bd{B}||^2,
\end{gather}
which becomes a function of \textcolor{black}{the $T \times (r+T)$ cross-covariance matrix between $\bd{Y}$ and $\bd{Z}$,} $\bd{S}_{YZ}=\frac{1}{N}\bd{Y}^{\top}\bd{Z}
=\left[
\frac{1}{N}\bd{Y}^{\top}\bd{F},
\frac{1}{N}\bd{Y}^{\top}\bd{U}
\right]
=
\left[
\bd{S}_{YF},
\bd{S}_{YU}
\right]
$ (Adachi \& Trendafilov, 2018, Theorem 2.1).
This expression indicates that only the second term $N \lVert \bd{S}_{YZ} - \bd{B} \rVert^{2}$ depends on $\bd{B}$, and that its solution can be obtained as 
\begin{gather}
\bd{\hat{\Lambda}} = \bd{S_{YF}},\hspace{3mm} \bd{\hat{D}} = \rm{diag}(\bd{S_{YU}}).\label{estB}
\end{gather}
Let $\bd{L}$ be the orthonormal matrix of eigenvectors and let $\bd{\Delta}$ be the diagonal matrix of (positive) eigenvalues obtained from the eigen-decomposition of
\begin{gather}
\bd{B}^\top \bd{S} \bd{B} = \bd{L} \bd{\Delta}^2 \bd{L}^\top, \label{eigendecomp}
\end{gather}
where $\bd{L}^\top \bd{L} = \bd{I}$. Adachi \& Trendafilov (2018, Lemma 3.1) shows that $\bd{S_{YZ}}$ can be calculated by 
\begin{gather}
\bd{S_{YZ}}=\bd{S}\bd{B}\bd{L}\bd{\Delta}^{-1}\bd{L}^\top. \label{estSYZ}
\end{gather}
As a result, in MDFA, given appropriate initial values and data \textcolor{black}{(sample covariance matrix of outcomes; \bd{S})}, one can obtain estimates of $\bd{B}$ by iteratively computing $\bd{S_{YZ}}$ (as a function of $\bd{B}$) by the Equation (\ref{estSYZ})
and updating $\bd{B}$ based on the computed $\bd{S_{YZ}}$ using the Equation (\ref{estB}) until the convergence criterion is met.

As a notable advantage of MDFA, it has been pointed out that it always provides proper solutions by exploiting eigenvalue decomposition (i.e., no Heywood cases; Yamashita, 2024; Terada, 2025). Moreover, theoretical developments have recently progressed, including extensions to high-dimensional settings and SEM that assumes structural relations among latent variables with $L_2$ penalization (Yamashita, 2024).

Note that the loss functions of MDFA (Equation~(\ref{lossMDFA})) and ULS 
(i.e., $(\bd{s}-\bd{\sigma}(\bd{\theta}))^\top(\bd{s}-\bd{\sigma}(\bd{\theta}))$, where $\bd{\Sigma}(\bd{\theta})=\bd{\Lambda}\bd{\Lambda}^{\top} + \bd{D}^{2}$) 
are not equivalent, and the estimates of $\bd{\theta}$ are not necessarily identical. In other words, MDFA performs fitting in the data space while treating $\bd{F}$ and $\bd{U}$ as additional parameters, whereas ULS in the SEM framework estimates the parameters from the perspective of covariance structure fitting. On the other hand, it has been empirically reported that the solutions obtained by MDFA are often very similar to the ML estimates in the context of factor analysis models (e.g., Adachi \& Trendafilov, 2018).

Although the statistical properties of the MDFA estimator had remained largely unexplored, Terada (2025) demonstrated that MDFA can be formulated not as a mere matrix-decomposition-based numerical procedure, but rather as a semiparametric ML estimator. Terada (2025) also showed that the loss function of MDFA coincides with the squared Bures--Wasserstein distance between $\bd{S}$ and $\bd{\Sigma}(\bd{\theta})$ (Proposition~3.1), and established both the consistency and asymptotic normality of the MDFA estimator (Theorem~3.4 and Theorem~3.5).
\subsection{Proposed approach}
One key idea of the proposed approach is to reinterpret the STARTS model within the framework of factor analysis, where the relations among latent variables are structured using model parameters. 
This perspective allows estimates of some parameters ($\phi^{2}$ and $\psi^{2}$) to be obtained in a manner analogous to MDFA, while the remaining parameters associated with within-person variability (\textcolor{black}{$\sigma_{w}^{2}$ and $\beta$}) are obtained by fitting the covariance structure corresponding to the auto-regression part of the model (which represents within-person variability; Equation (\ref{wcov})) to a covariance matrix as a function of the (submatrix of) $\bd{S_{YZ}}$, which is calculated in the MDFA step. The details of the procedure are presented below.
\subsubsection{The STARTS Model as an MDFA representation}
From Equations (\ref{withinexpand2}) and (\ref{starts1}), it follows that $\textcolor{black}{y_{iT}}$ can be expressed as a total of $(T+2)$ terms related to $s_{i}$, $y^*_{i1}$, $d_{i2},\dots,d_{iT}$ and $\epsilon_{it}$. With this point in mind, we explain how the STARTS model corresponds to the MDFA representation and how its estimation can be carried out.

Once again, let $\bd{Y}$ denote an $N \times T$ matrix of outcomes. Then,
Equation (\ref{starts1}) can now be expressed as
\begin{gather}
\bd{Y}=\bd{s}\bd{1_T}^\top+\bd{Y^*}+\bd{E},\label{startsmat}
\end{gather}
where $\bd{s}=(s_{1},\dots,s_{N})^\top$, $\bd{Y^*}=(\bd{y^*_1},\dots,\bd{y^*_i},\dots,\bd{y^*_N})^\top$, where $\bd{y^*_i}=(y^*_{i1},\dots,y^*_{iT})^\top$, is an $N\times T$ matrix of within-person variability \textcolor{black}{terms}, and $\bd{E}=(\bd{e_1},\dots,\bd{e_i},\dots,\bd{e_N})^\top$, where $\bd{e_i}=(\epsilon_{i1},\dots,\epsilon_{iT})^\top$, is an $N\times T$ matrix of measurement error \textcolor{black}{terms}.

From the relation (\ref{reg}), $\bd{Y^*}$ can be expressed as
\begin{align}
\bd{Y^*}=\bd{Y^*}\bd{\Gamma}+\bd{R}, \label{withinmat}
\end{align}
where \bd{\Gamma} denotes a $T\times T$ coefficient matrix that includes $\beta$:
\begin{align}
\bd{\Gamma}=\left(
\begin{array}{cc}
\bd{0_{T-1}} & \beta\bd{I_{T-1}} \\
0 & \bd{0_{T-1}}^\top 
\end{array}
\right)
.\end{align}
$\bd{R}=(\bd{r_1},\dots,\bd{r_i},\dots,\bd{r_N})^\top$, where
$\bd{r_i}=(d_{i1},d_{i2},\dots,d_{iT})^\top$, is an $N\times T$ matrix for residuals that are mutually uncorrelated. Note that $y^*_{i1}$ is an exogeneous variable and then the first column of Equation ($\ref{withinmat}$) shows the relation $d_{i1}=y^*_{i1}$ (i.e., $\bd{r_i}=(y^*_{i1},d_{i2},\dots,d_{iT})^\top$).

Assuming that inverse matrix of $(\bd{I_T}-\bd{\Gamma})$ exists, from Equation (\ref{withinmat}) \bd{Y^*} can be reexpressed as
\begin{align}
\bd{Y^*}=\bd{R}(\bd{I_T}-\bd{\Gamma})^{-1}.
\end{align}
By substituting this result into the Equation (\ref{startsmat}), we obtain
\begin{align}
\bd{Y}=\bd{s}\bd{1_T}^\top+\bd{R}(\bd{I_T}-\bd{\Gamma})^{-1}+\bd{E}.
\end{align}
For estimation, we now reparameterize the factor score (including residuals and measurement errors) matrices as
\begin{gather}
\bd{s}=\bd{\tilde{s}}\phi,\hspace{3mm} \bd{R}=\tilde{\bd{R}}\mathrm{diag}(\textcolor{black}{\sigma_{w},\sqrt{1-\beta^2}\,\sigma_w\bd{1_{T-1}}}),\hspace{3mm} \bd{E}=\bd{\tilde{U}}\psi\bd{I_T}=\bd{\tilde{U}}\bd{\tilde{D}},
\end{gather}
where $\frac{1}{N}\bd{\tilde{s}}^\top\bd{\tilde{s}}=1$, $\frac{1}{N}\tilde{\bd{R}}^\top\tilde{\bd{R}}$=$\bd{I_T}$, $\frac{1}{N}\bd{\tilde{U}}^\top\bd{\tilde{U}}$=$\bd{I_T}$, and $\bd{\tilde{D}}=\psi\bd{I_T}$. $\mathrm{diag}(\textcolor{black}{\sigma_{w},\sqrt{1-\beta^2}\,\sigma_w\bd{1_{T-1}}})$ denotes the diagonal matrix whose diagonal consists of $\textcolor{black}{\sigma_{w}}$ followed by $T-1$ repetitions of $\textcolor{black}{\sqrt{1-\beta^2}\,\sigma_w}$. Then $\bd{Y}$ becomes
\begin{align}
\bd{Y}=\bd{\tilde{s}}\phi\bd{1_T}^\top+\tilde{\bd{R}} \mathrm{diag}(\textcolor{black}{\sigma_{w},\sqrt{1-\beta^2}\,\sigma_w\bd{1_{T-1}}})(\bd{I_T}-\bd{\Gamma})^{-1}+\bd{\tilde{U}}\bd{\tilde{D}}.
\end{align}
\bd{Y} can be further expressed as
\begin{align}
\bd{Y}=\bd{\tilde{F}}\bd{\tilde{\Lambda}}^\top+\bd{\tilde{U}}\bd{\tilde{D}},
\end{align}
where $\bd{\tilde{F}}=(\bd{\tilde{s}},\tilde{\bd{R}})$, 
$\bd{\tilde{\Lambda}}=[\bd{\lambda_{\tilde{s}}},\bd{\Lambda_{\tilde{R}}}]=[\phi\bd{1_T},\mathrm{diag}(\textcolor{black}{\sigma_{w},\sqrt{1-\beta^2}\,\sigma_w\bd{1_{T-1}}})(\bd{I_T}-\bd{\Gamma})^{-1}]$. From this, it follows that the STARTS model can be reformulated in a manner consistent with the MDFA representation (in the case where the relations among common factors are structured by model parameters) by introducing $\bd{\tilde{F}}$, which consists of $r=T+1$ common factors—namely, one trait factor $\bd{\tilde{s}}$, the initial within-person variability $y^{*}_{i1}$ and $(T-1)$ residuals within $\tilde{\bd{R}}$—, and $\bd{\tilde{U}}$, which represents the measurement error and is now treated as the unique factor matrix. Moreover, in this formulation, $\bd{\tilde{\Lambda}}$ contains information relating to variance and covariance for the trait factor and the within-person variability, rather than factor loadings as in the standard factor analysis model.
\subsubsection{Two-stage parameter estimation using MDFA approach}
By considering the Frobenius-norm-based loss function corresponding to Equation~(\ref{lossMDFA}), solutions of $\bd{\tilde{\Lambda}}$ and $\bd{\tilde{D}}$ can be obtained analogously to Equation~(\ref{estB}) as:  
\begin{align}
\bd{\hat{\tilde{\Lambda}}} = \bd{S_{Y\tilde{F}}} = [\,\bd{s_{Y\tilde{s}}},\,\bd{S_{Y\tilde{R}}}\,], \hspace{3mm}
\bd{\hat{\tilde{D}}} = \mathrm{diag}(\bd{S_{Y\tilde{U}}}).
\end{align}
\bd{s_{Y\tilde{s}}} ($T \times 1$) is a vector that contains information about the variance of the trait factor, whereas \bd{S_{Y\tilde{R}}} ($T \times T$) is a matrix that contains information about the within-person variance--covariance structure. The parameters $\phi^2$ and $\psi^2$ are included in \bd{\tilde{\Lambda}} (i.e., \bd{\lambda_{\tilde{s}}}) and \bd{\tilde{D}} respectively, without being confounded with other parameters. Therefore, by using arithmetic means, we obtain
\begin{align}
\hat{\phi}^2 &= \left(\frac{1}{T}\bd{1}^{\top}_T \bd{s_{Y\tilde{s}}}\right)^2, \hspace{3mm}
\hat{\psi}^2 = \left(\frac{1}{T}\mathrm{tr}(\bd{S_{Y\tilde{U}}})\right)^2. \label{estp}
\end{align}
In the original MDFA framework, it is formulated based on the standard factor analysis model with the variance of the common factor fixed to one, and $\bd{\Lambda}$ contains factor loading information. In contrast, within the proposed framework for the STARTS model, the coefficients (factor loadings) associated with the $(T+1)$ common factors are all fixed to one. Consequently, it should be noted that each element in $\bd{s}_{Y\tilde{s}}$, for example, contains information for the standard deviation of the stable trait factor.

Let $\bd{\Sigma}^*(\bd{\theta_1})$, where $\bd{\theta_1}=(\textcolor{black}{\sigma_w^2, \beta})^\top$, denote the variance–covariance structure at the within-person variability, and partition the full parameter vector $\bd{\theta}$ as $\bd{\theta}=(\bd{\theta_1}^\top,\bd{\theta_2}^\top)^\top$, where $\bd{\theta_2}=(\phi^2, \psi^2)^\top$. Given the relation of 
\begin{align}
\bd{\Sigma^*}(\bd{\theta_1})&=\bd{\Lambda_{\tilde{R}}}{\bd{\Lambda_{\tilde{R}}}}^\top \notag \\
&=(\bd{I_T}-\bd{\Gamma})^{-1}\mathrm{diag}(\textcolor{black}{\sigma_{w},\sqrt{1-\beta^2}\,\sigma_w\bd{1_{T-1}}})
\mathrm{diag}(\textcolor{black}{\sigma_{w},\sqrt{1-\beta^2}\,\sigma_w\bd{1_{T-1}}})(\bd{I_T}-\bd{\Gamma})^{-1}\notag \\
&=(\bd{I_T}-\bd{\Gamma})^{-1}\mathrm{diag}(\textcolor{black}{\sigma^2_{w},(1-\beta^2)\sigma_w^2\bd{1_{T-1}}})(\bd{I_T}-\bd{\Gamma})^{-1},
\end{align}
by computing
\begin{align}
\bd{S^*} &= \bd{S^{\triangle}_{Y\tilde{R}}}{\bd{S^{\triangle}_{Y\tilde{R}}}}^\top,
\end{align}
where $\bd{S^{\triangle}_{Y\tilde{R}}}$ denotes the upper triangular part of $\bd{S_{Y\tilde{R}}}$ (note that $y_{it}$ and $d_{it'}$ $(t'>t)$ is assumed to be independent in the STARTS model; Figure~1), we can obtain the solution for the variance–covariance structure of the within-person variability $y^*_{it}$ that appears in Equations (\ref{varst})–(\ref{covst}). Namely,
\begin{equation}
\bd{S}^* = \{ s_{tt'}^* \} = \{ \widehat{cov}(y_{it}^*,\, y_{it'}^*) \}.
\end{equation}
As shown in Equation (\ref{wcov}), in the STARTS framework variances and covariances for the within-person variability are structured as functions of $\bd{\theta_1}$, and therefore, unlike Equation (\ref{estp}), the solution for $\bd{\theta_1}$ cannot be obtained directly. Thus, some method is required to approximate the variance–covariance structure at the within-person variability 
$\bd{\Sigma}^*(\bd{\theta_1})$ to $\bd{S^*}$. Here, following a Frobenius norm-based criterion in the sense that no distributional assumptions are imposed, we obtain solutions for $\bd{\theta_1}$ based on the SEM framework using the ULS criterion:
\begin{gather}
(\bd{s^*}-\bd{\sigma}^*(\bd{\theta_1}))^\top(\bd{s^*}-\bd{\sigma}^*(\bd{\theta_1})),\label{MDFAULS}
\end{gather}
where $\bd{s^*}=vech(\bd{S^*})$ and $\bd{\sigma}^*(\bd{\theta_1})=vech(\bd{\Sigma^*}(\bd{\theta_1}))$.

From the preceding discussion, it becomes clear that the proposed method adopts a two-stage estimation procedure in which the model parameters are divided into two subsets, $\bd{\theta}_1$ and $\bd{\theta}_2$, and updated
separately, because $\bd{\Sigma}^*(\bd{\theta}_1)$ is structured as a function of $\bd{\theta}_1$ in the STARTS model. This point differs from the original MDFA framework. As a consequence, even if the parameter updates are iterated in the same manner as in the MDFA algorithm, the loss function:
\begin{gather}
\bigl\|\bd{Y} - 
(\tilde{\bd{F}}\tilde{\bd{\Lambda}}^{\top}
+ \tilde{\bd{U}}\tilde{\bd{D}}) \bigr\|^2,
\end{gather}
or its equivalent and more succinct representation:
\begin{gather}
T_{\mathrm{MDFA}}(\bd{\theta})=N\|\bd{S_{Y\tilde{Z}}} - \bd{\tilde{B}}\|^2, \label{MDFAloss}
\end{gather}
where $\bd{S_{Y\tilde{Z}}} = [\bd{S_{Y\tilde{F}}}, \bd{S_{Y\tilde{U}}}]$ (by using the relation $\bd{S}=\bd{S_{YZ}}\bd{S_{YZ}}^\top$; Adachi \& Trendafilov, 2018, Lemma 4.1) and $\bd{\tilde{B}} = (\bd{\tilde{\Lambda}}^{\top}, \bd{\tilde{D}})$, does not necessarily decrease monotonically. Furthermore, once the function value begins to increase, the decrease does not resume in many cases, and a tendency was observed in which certain variance components approach the boundary value
of zero.\footnote{In the preliminary stage of this study, we confirmed that this trend was not limited to the use of $T_{\mathrm{MDFA}}(\bd{\theta})$. A similar pattern was also observed even when an alternative ULS-based criterion corresponding to Equation~(\ref{ULS}), or its weighted version incorporating information for variances of each variable, was used instead.}

Based on these considerations, we employed an early-stopping rule in which the iteration terminated if the loss function failed to improve for 10 consecutive steps (i.e., patience = 10). In addition, the optimization was performed using multiple initial starting values like Yamashita (2024), since the loss function in MDFA is intrinsically non-convex, stemming from the presence of two unknown matrices ($\bd{\tilde{Z}}$ and $\bd{\tilde{B}}$) that appears jointly, and is therefore characterized by a landscape with many local optima. The final estimate is chosen as the solution that yields the smallest function value in Equation (\ref{MDFAloss}) among all attempts.

Summarizing the above explanations, the proposed two-stage estimation algorithm obtains the estimate of $\bd{\theta}$ in the STARTS model through the following steps:

\begin{enumerate}
\item[0.] 
Prepare $M$ distinct initial values for the parameter vector 
$\bd{\theta}$, denoted as 
$\{\bd{\theta}^{(0,m)}\}_{m=1}^M$.

\item[1.]
Select the $m$-th initial value $\bd{\theta}^{(0,m)}$ and 
use it as the starting point of the iterative updates.

\item[2.]
Update 
$\bd{S_{Y\tilde{Z}}}$ using $\tilde{\bd{B}}$ (that 
is a function of \bd{\theta}) and the relations in Equations (\ref{eigendecomp})-(\ref{estSYZ}) as 
\begin{gather}
\bd{S_{Y\tilde{Z}}}=\bd{S}\tilde{\bd{B}}\bd{\tilde{L}}\bd{\tilde{\Delta}}^{-1}\bd{\tilde{L}}.
\end{gather}
Here, $\bd{\tilde{L}}$ and $\bd{\tilde{\Delta}}$ are calculated from the eigen-decomposition of
\begin{gather}
\tilde{\bd{B}}^\top \bd{S}\tilde{\bd{B}} = \bd{\tilde{L}} \bd{\tilde{\Delta}}^2 \bd{\tilde{L}}^\top.
\end{gather}

\item[3.1.]
Using the updated $\bd{S_{Y\tilde{Z}}}$, update the subvector 
$\bd{\theta_1}$ by minimizing the ULS criterion in Equation~(\ref{MDFAULS}).

\item[3.2.]
Using the same $\bd{S_{Y\tilde{Z}}}$, update the remaining 
subvector $\bd{\theta_2}$ based on Equation~(\ref{estp}).

\item[4.]
Repeat Steps~2 and 3 until convergence criterion (the change in each parameter before and after the update is less than $\varepsilon \,(=10^{-6})$) is met or the loss function $T_{\mathrm{MDFA}}(\bd{\theta})$ failed to improve for 10 consecutive steps (i.e., patience = 10). 

\item[5.]
Repeat Steps~1--4 for all $m=1,\dots,M$. 
Among the $M$ resulting estimates, select the one that yields the smallest value of the loss function $T_{\mathrm{MDFA}}(\bd{\theta})$ and adopt it as the final estimates.
\end{enumerate}\textcolor{black}{
\subsubsection{Standard errors of estimates}
In the original proposal of MDFA, the calculation of standard errors for the parameter estimates was not directly discussed. In this paper, standard errors for the parameter estimates are obtained by treating the converged TS-MDFA estimator as the solution to a sample fixed-point equation. The asymptotic linearization is based on the fixed-point argument of Dominitz and Sherman (2005, Theorem 4 and its proof), together with a first-order implicit-function and delta-method expansion and the standard sandwich theory for estimating equations (e.g., Stefanski \& Boos, 2002). The asymptotic covariance matrix of the sample covariance statistics is estimated using an empirical fourth-moment estimator, as in general covariance structure theory (e.g., Browne, 1984).}

\textcolor{black}{Define $\mathcal{T}_{\mathrm{TS}}(\bd{\theta};\bd{S})$ as one complete updating step in the proposed two-stage ($\mathrm{TS}$) approach applied to $\bd{\theta}$ and the sample covariance matrix $\bd{S}$. Because the converged estimate $\hat{\bd{\theta}}$ is expected to satisfy $\hat{\bd{\theta}} = \mathcal{T}_{\mathrm{TS}}(\hat{\bd{\theta}};\bd{S})$ (i.e., it remains almost unchanged after applying the update), it approximately satisfies
\begin{equation}
  \bd{U}_{\mathrm{TS}}(\hat{\bd{\theta}},\bd{S})=\bd{0},
  \qquad
  \bd{U}_{\mathrm{TS}}(\bd{\theta},\bd{S})
  =
  \bd{\theta}
  -
  \mathcal{T}_{\mathrm{TS}}(\bd{\theta};\bd{S}).
  \label{eq:tsmdfa_fixed_point_se}
\end{equation}
Let $\bd{s}=vech(\bd{S})$ and 
\begin{gather}
  \bd{G}_{\theta,\mathrm{TS}}
  =
  \left.
  \frac{\partial \bd{U}_{\mathrm{TS}}(\bd{\theta},\bd{s})}
       {\partial \bd{\theta}^\top}
  \right|_{(\hat{\bd{\theta}},\bd{s})},
  \qquad
  \bd{G}_{s,\mathrm{TS}}
  =
  \left.
  \frac{\partial \bd{U}_{\mathrm{TS}}(\bd{\theta},\bd{s})}
       {\partial \bd{s}^\top}
  \right|_{(\hat{\bd{\theta}},\bd{s})},
\end{gather}
and these derivative matrices can be evaluated numerically by central finite differences. To estimate the sampling variability of $\bd{s}$, define
$
  \bd{q}_i
  =
  vech\left((\bd{y}_i-\bar{\bd{y}})(\bd{y}_i-\bar{\bd{y}})^\top\right)$,
where $\bar{\bd{y}}=N^{-1}\sum_{i=1}^N \bd{y}_i$. Let
\begin{gather}
  \widehat{\bd{\Gamma}}_s
  =
  N^{-1}\sum_{i=1}^N
  (\bd{q}_i-\bar{\bd{q}})(\bd{q}_i-\bar{\bd{q}})^\top,
  \qquad
  \bar{\bd{q}}=N^{-1}\sum_{i=1}^N \bd{q}_i .
\end{gather}
Then, by a first-order Taylor expansion of Equation~\eqref{eq:tsmdfa_fixed_point_se},
covariance matrix of $\hat{\bd{\theta}}$ can be estimated by the following
sandwich-type estimator:
\begin{equation}
  \widehat{Var}[\hat{\bd{\theta}}]
  =
  \bd{D}_{\mathrm{TS}}
  \widehat{\bd{\Gamma}}_s
  \bd{D}_{\mathrm{TS}}^\top/N,
  \qquad
  \bd{D}_{\mathrm{TS}}
  =
  -\bd{G}_{\theta,\mathrm{TS}}^{-1}\bd{G}_{s,\mathrm{TS}} .
  \label{eq:tsmdfa_sandwich_se}
\end{equation}
Standard errors can be obtained as the square roots of the diagonal elements of $\widehat{Var}[\hat{\bd{\theta}}]$.
}
\section{Simulation}
\subsection{Method}
The primary objective of this simulation study is to evaluate the performance of the proposed two-stage estimation (hereafter, we refer to this as TS-MDFA) for the STARTS model. Specifically, we assess the frequency of improper solutions, bias, RMSE \textcolor{black}{and coverage} of estimates, as well as computation time under various conditions, and compare the results with those obtained using ML, CML and ULS estimation in SEM, as well as Bayesian estimation.

The simulation design varied the number of time points as $T = (4, 6, 8)$, the sample size as $N = (200, 1000)$, the variance of measurement error as \textcolor{black}{$\psi^2 = (0.1, 0.3)$, and the autoregressive coefficient as $\beta = (0.3, 0.5)$}. The variance of the 
\textcolor{black}{outcome $y_{it}$ was fixed at
$Var(y_{it}) = 1.0$. The ratio of the variance of the
within-person variability ($\sigma_w^2$), to the variance of the stable trait
factor ($\phi^2$), was fixed at $1:1$, such that
$\sigma_w^2 = \phi^2 = 0.5(1-\psi^2)$. Then, using the relation in
Equation~(\ref{within}), the residual variance $\omega^2$ was specified based on the
$\sigma_w^2$ and $\beta$.} These settings yielded a fully factorial design with $3 \times 2 \times 2$ $\textcolor{black}{ \times 2 = 24}$ conditions, and \textcolor{black}{500} replications were conducted per condition. 

For each condition, the model-implied covariance matrix was computed based on the true parameter values, and data following the STARTS model were generated under the assumption of multivariate normality. ML, CML and ULS estimations were performed using the \texttt{lavaan} package (version 0.6-20; Rosseel, 2012) in R (Version 4.5.1; R Core Team, 2025), whereas Bayesian estimation was carried out using \textcolor{black}{a custom-written \textsf{Stan} model implemented through the \texttt{rstan} package}. \textcolor{black}{In CML, optimization was performed under constraints requiring all variance parameters to be positive. In \texttt{lavaan}, such inequality constraints are internally treated as lower bounds on the parameters. When nonlinear constraints such as that in Equation~(\ref{within}) are imposed, the constrained optimization routine \texttt{NLMINB.CONSTR} is used.}

\textcolor{black}{Following L$\ddot{\mathrm{u}}$dtke et al.\ (2018), prior distributions were specified for the four free parameters in the Bayesian analysis. For the variance parameters, the prior distribution was specified as $\mathrm{Inv\text{-}Gamma}(v_0/2,\,v_0\sigma_0^2/2)$, with $v_0=3$ and $\sigma_0^2=1/3$. Accordingly, the priors were $\phi^2 \sim \mathrm{Inv\text{-}Gamma}(1.5,0.5)$, $\psi^2 \sim \mathrm{Inv\text{-}Gamma}(1.5,0.5)$, and $\sigma_w^2 \sim \mathrm{Inv\text{-}Gamma}(1.5,0.5)$, where the second parameter denotes the scale parameter. The prior distribution of the autoregressive coefficient was specified as $\beta \sim \mathrm{Beta}((v_0+1)\beta_0,\,(v_0+1)(1-\beta_0))$, with $\beta_0=0.5$, resulting in $\beta \sim \mathrm{Beta}(2,2)$.} 

Posterior sampling was performed using the No-U-Turn Sampler (NUTS), an adaptive variant of Hamiltonian Monte Carlo (HMC) implemented in \textsf{Stan}. NUTS adaptively tunes step sizes and trajectory lengths to achieve efficient exploration of the posterior \textcolor{black}{without requiring users to directly specify the step size or trajectory length},
\textcolor{black}{and can provide} higher effective sample sizes (a measure of the amount of independent information contained in an autocorrelated Markov chain) per unit time compared with traditional Gibbs sampling or Metropolis--Hastings algorithms.
In light of potential non-convergence issues that may arise in the structurally intricate STARTS model, the warm-up phase was set to 1,000 iterations, followed by 10,000 sampling iterations, using two independent Markov chains. Convergence was evaluated using the potential scale reduction statistic $\hat{R}$. The posterior means were used as the parameter estimates. \textcolor{black}{Posterior standard deviations were used as standard error estimates, and the 2.5th and 97.5th percentiles of the posterior draws were used to construct 95\% credible intervals.}

For comparability and to mitigate potential problem of convergence to local optima—we generated \textcolor{black}{$M=10$} sets of initial values in each condition \textcolor{black}{for TS-MDFA. In TS-MDFA, the model was estimated using 10 sets of initial values, whereas for ML, CML, and ULS, the model was estimated using only the first set from this common pool.} Initial values were sampled from \textcolor{black}{$\phi^2 \sim \mathrm{Gamma}(3,6)$,
$\psi^2 \sim \mathrm{Gamma}(2,5)$, 
$\sigma_w^2 \sim \mathrm{Gamma}(4,4)$, and $\beta\sim\mathrm{Beta}(4,4)$.} For each estimation method, estimates were counted as improper if at least one of the estimates of variance parameters $(\phi^2, \psi^2, \textcolor{black}{\sigma_{w}^2}, \omega^2)$ was less than $0.0001$.

Following Siepe et al.\ (2024), here we summarize the computational aspects of the simulation study in greater detail. All analyses were executed on a system running Windows 10 x64 (build 19045), using \texttt{R} Version 4.5.1 on the x86$\_$64-w64-mingw32/x64 platform with packages: the \texttt{lavaan} and \texttt{blavaan} packages to fit structures implied by the STARTS model, the rnorm( ) function included in the stats package (Version 4.5.1; R Core Team) to generate data, the eigen( ) function included in the base package (Version 4.5.1; R Core Team) for eigen-decomposition, the rstan package (Version 2.32.7; Stan Development Team, 2024) for Bayesian estimation via HMC, and the ggplot2 package (Version 4.0.0; Wickham, 2016) to create visualizations. \textcolor{black}{ChatGPT (GPT-5.5) was used to assist in developing some of the helper functions in the simulation code.} A full \texttt{sessionInfo()} output with additional computational details, and all code is available at the Open Science Framework (\textcolor{black}{\url{https://osf.io/4ungy/overview?view_only=7fa8a1ceb55c4fa6828e2f2dcef8beaf}}).

\subsection{Result}
Across all conditions, the proportion of cases in which the potential scale reduction statistics $\hat{R}$ for all parameters were below \textcolor{black}{1.05} under Bayesian estimation was \textcolor{black}{99.9 \%} on average. Notably, larger values of $\hat{R}$ tended to be observed when the sample size was large ($N = 1000$). \textcolor{black}{Accordingly, all subsequent analyses were basically conducted using the full set of cases.}

\subsubsection{Frequency of improper solutions}
\begin{table}[htbp]
\textcolor{black}{
\centering
\caption{Proportions of improper solutions across estimation methods ($\beta = 0.3$).}
\label{tab:improper_beta_0p3}
\begin{tabular}{rrrrrrrrrr}
\toprule
$T$ & $N$ & $\psi^2$ & ML & ULS & CML & ALL & TS-MDFA & BAY & TS-MDFA($10^{-2}$) \\
\midrule
4 & 200 & 0.1 & 0.64 & 0.69 & 0.56 & 0.47 & 0.00 & 0.00 & 0.19 \\
4 & 200 & 0.3 & 0.67 & 0.75 & 0.61 & 0.49 & 0.00 & 0.00 & 0.15 \\
4 & 1000 & 0.1 & 0.38 & 0.44 & 0.35 & 0.28 & 0.00 & 0.00 & 0.07 \\
4 & 1000 & 0.3 & 0.41 & 0.47 & 0.39 & 0.30 & 0.00 & 0.00 & 0.01 \\
6 & 200 & 0.1 & 0.41 & 0.47 & 0.33 & 0.23 & 0.06 & 0.00 & 0.33 \\
6 & 200 & 0.3 & 0.39 & 0.50 & 0.33 & 0.25 & 0.01 & 0.00 & 0.24 \\
6 & 1000 & 0.1 & 0.28 & 0.32 & 0.22 & 0.13 & 0.00 & 0.00 & 0.09 \\
6 & 1000 & 0.3 & 0.19 & 0.24 & 0.16 & 0.11 & 0.00 & 0.00 & 0.01 \\
8 & 200 & 0.1 & 0.39 & 0.44 & 0.32 & 0.21 & 0.11 & 0.00 & 0.39 \\
8 & 200 & 0.3 & 0.31 & 0.41 & 0.25 & 0.17 & 0.02 & 0.00 & 0.26 \\
8 & 1000 & 0.1 & 0.22 & 0.29 & 0.18 & 0.10 & 0.00 & 0.00 & 0.16 \\
8 & 1000 & 0.3 & 0.11 & 0.12 & 0.07 & 0.04 & 0.00 & 0.00 & 0.01 \\
\bottomrule
\end{tabular}
\begin{flushleft}
\footnotesize Note. Improper solutions are defined as cases in which at least one variance-parameter estimate is less than 0.0001. ALL indicates the proportion of replications in which ML, ULS, and CML all yielded improper solutions. TS-MDFA($10^{-2}$) indicates the corresponding proportion for TS-MDFA when the threshold is changed to 0.01.
\end{flushleft}
}
\end{table}
Table~1 presents, for each estimation method, the frequency of improper solutions under each condition \textcolor{black}{when $\beta=0.3$.} First, we focus on ML, CML and ULS, which correspond to the first three columns. \textcolor{black}{For each of these estimation methods, improper solutions occurred in at least some replications under every condition, with ULS exhibiting the highest overall frequency. Although the proportion of improper solutions for ML was lower than that for ULS, it was approximately 2\% to 8\% higher than that for CML across the conditions.}
For example, under the condition $T = 4$, $N = 200$, and $\psi^2 = 0.3$, the proportions of improper solutions were \textcolor{black}{67\%} for ML, \textcolor{black}{61\%} for CML, and \textcolor{black}{75\%} for ULS. 

Moreover, the proportion in which all three methods resulted in improper solutions reached as high as \textcolor{black}{49\%}. Improper solutions were most frequently observed in the estimates of $\psi^2$, and this tendency became more pronounced as $T$ and $N$ increased. These results underscore the severity of the improper-solution problem when applying the STARTS model. \textcolor{black}{Moreover, as observed in the preceding results, although CML yielded a slightly lower frequency of improper solutions than ML, in many cases at least one variance parameter—most commonly $\psi^2$—was estimated at the boundary value of zero.}

Both $N$ and $T$ were associated with reductions in the frequency of improper solutions as they increased. However, the effect of $T$ appeared to be relatively limited when increasing from $T = 6$ to $T = 8$, whereas the impact of $N$ was more pronounced overall. In addition, differences attributable to $\psi^2$ were generally small unless $T=8$. \textcolor{black}{Regarding $\beta$, the overall proportion of improper solutions tended to be lower under $\beta=0.5$ (see Table~S1 in the Online Supplemental Material), presumably because the respective roles of $\sigma_w^2$ and $\psi^2$ in representing the covariance structure were more clearly differentiated. Nevertheless, the differences across estimation methods and conditions followed patterns similar to those observed when $\beta=0.3$ (Table 1).}

Notably, even under conditions such as $T = 8$, which is relatively large in contexts where STARTS models are typically applied, improper solutions still occurred at proportions of approximately 10\% to \textcolor{black}{40\%} in at least one estimation method \textcolor{black}{when $\beta=0.3$}. Furthermore, except for the condition with $N = 1000$ and $ \textcolor{black}{\psi^2 = 0.3}$, the proportion in which all three methods (ML, CML and ULS) yielded improper solutions remained around  \textcolor{black}{10\%--20\%}. Consistent with these findings, Usami et al. (2019) reported high frequencies of improper solutions for the bivariate STARTS model. The present simulation further demonstrates that, even in the univariate case---where risks such as non--positive-definite covariance matrices among multiple trait factors do not arise---improper solutions can occur at high frequencies in ML, CML and ULS.

\textcolor{black}{On the other hand, no improper solutions—defined as cases in which at least one variance parameter estimate fell below 0.0001—were observed for Bayesian estimation. Similarly, TS-MDFA yielded no improper solutions except under conditions in which both $N$ and $\psi^2$ were small.} Namely, it can be seen that TS-MDFA effectively avoids improper solutions due to its parameter-updating procedure based on eigenvalue decomposition, while Bayes avoids them due to the specification of prior distributions.

However, it should be noted that although TS-MDFA does not exhibit the same tendency as CML to consistently yield estimates that are stuck at the boundary value of zero, it nevertheless shows a non-negligible tendency to produce relatively small estimates. As a more lenient criterion, when improper solutions are provisionally defined as cases in which the estimate of at least one variance parameter was less than 0.01, TS-MDFA yields improper solutions in approximately \textcolor{black}{5\%--30\% of cases except when $\psi^2=0.3$ and $N=1000$}, although this proportion is smaller overall than that observed in ML, CML and ULS (see the rightmost column of Table~1). This proportion tends to be higher when $N$ is small ($N=200$). Furthermore, as with the other estimation methods, most of improper solutions in TS-MDFA were observed in the estimates of $\psi^2$, and this tendency became more pronounced as $T$ and $N$ increased.
\subsubsection{Computation time}

\begin{table}[htbp]
\textcolor{black}{
\centering
\caption{Required computation time (in seconds) across estimation methods ($\beta = 0.3$).}
\label{tab:computation_time_beta_0p3}
\begin{tabular}{ccc|rrrrrrrrrr}
\hline
\multicolumn{3}{c|}{} &
\multicolumn{2}{c}{ML} &
\multicolumn{2}{c}{CML} &
\multicolumn{2}{c}{Bayes} &
\multicolumn{2}{c}{ULS} &
\multicolumn{2}{c}{TS-MDFA} \\
\hline
$T$ & $N$ & $\psi^2$ &
$M$ & $SD$ &
$M$ & $SD$ &
$M$ & $SD$ &
$M$ & $SD$ &
$M$ & $SD$ \\
\hline
4 & 200 & 0.1 & 8.2 & 8.5 & 6.9 & 7.4 & 15.6 & 2.6 & 8.2 & 8.5 & 3.7 & 3.5 \\
 &  & 0.3 & 15.3 & 15.9 & 13.6 & 16.8 & 19.8 & 2.2 & 12.0 & 12.0 & 6.6 & 6.4 \\
 & 1000 & 0.1 & 5.9 & 6.9 & 6.8 & 8.1 & 55.4 & 4.8 & 4.1 & 5.3 & 4.5 & 3.7 \\
 &  & 0.3 & 7.4 & 8.8 & 7.6 & 9.3 & 61.1 & 7.8 & 4.0 & 3.8 & 4.3 & 3.4 \\
\hline
6 & 200 & 0.1 & 18.0 & 21.0 & 20.5 & 23.6 & 18.6 & 1.8 & 11.0 & 11.0 & 4.8 & 4.4 \\
 &  & 0.3 & 27.4 & 32.1 & 27.8 & 34.4 & 22.2 & 2.9 & 16.6 & 15.6 & 6.7 & 6.4 \\
 & 1000 & 0.1 & 13.6 & 17.2 & 15.9 & 19.1 & 71.6 & 6.1 & 6.9 & 5.4 & 4.9 & 3.4 \\
 &  & 0.3 & 15.1 & 18.3 & 17.9 & 21.5 & 88.0 & 12.8 & 7.0 & 5.3 & 5.2 & 3.4 \\
\hline
8 & 200 & 0.1 & 23.9 & 26.1 & 26.0 & 26.9 & 20.6 & 1.3 & 16.6 & 14.2 & 4.6 & 3.5 \\
 &  & 0.3 & 25.2 & 27.1 & 24.0 & 25.3 & 22.5 & 2.0 & 19.9 & 17.6 & 4.2 & 3.8 \\
 & 1000 & 0.1 & 25.0 & 28.9 & 26.4 & 29.5 & 91.2 & 15.5 & 14.2 & 10.2 & 6.2 & 4.0 \\
 &  & 0.3 & 35.0 & 42.7 & 38.5 & 47.3 & 142.6 & 30.8 & 19.0 & 13.1 & 8.4 & 5.9 \\
\hline
\end{tabular}
\begin{flushleft}
\footnotesize
\textit{Note.} Result for TS-MDFA is based on 10 multiple starts. Results for Bayes are based on 11,000 sampling iterations for each of two independent Markov chains. $M$ = mean; $SD$ = standard deviation.
\end{flushleft}
}
\end{table}
Table~2 presents the means (standard deviations) of the computation times required to obtain parameter estimates across \textcolor{black}{500} replications for each condition and each of the five estimation methods. The computation times reported for \textcolor{black}{TS-MDFA} correspond to the time required to obtain the final estimates under $M = \textcolor{black}{10}$ sets of initial values. For the Bayesian estimation, the computation times represent the time required to run 11,000 sampling iterations for each of two independent Markov chains.

As shown in Table~2, \textcolor{black}{TS-MDFA} was the fastest method overall, \textcolor{black}{and it} typically completed the whole estimation within approximately 10 seconds in each condition. \textcolor{black}{Although ULS was the second-fastest method, it required approximately two to more than four times as much computation time as TS-MDFA, particularly under conditions with the smaller sample size of $N=200$. ML and CML were, on average, slower than ULS, and the differences became more pronounced as $T$ and $N$ increased, with computation times differing by as much as approximately a factor of two.}

Among the five methods, \textcolor{black}{Bayesian estimation was} the most computationally demanding. In particular, the computation time for this method increased substantially as $N$ becomes larger. When \textcolor{black}{$N = 1,000$,  Bayesian estimation required on average approximately one to two minutes}. 

\textcolor{black}{The relatively long computation times observed for ML, CML, and ULS may be partly attributable to the STARTS model setting, in which estimation must be performed subject to the constraint specified in Equation~(\ref{within}). In addition, ML and CML exhibited relatively large standard deviations in computation time, indicating that their computational costs can vary considerably depending on the characteristics of the observed data.}

\textcolor{black}{Trends similar to those described above were also observed when $\beta=0.5$ (Table~S2 in the Online Supplemental Material)}. When the comparison was restricted to cases in which no improper solutions occurred for each method, computation times tended to be shorter overall while the relative patterns observed across methods remained consistent with those described above (\textcolor{black}{see Tables~S3 and S4 in the Online Supplemental Material)}.

\textcolor{black}{It should be noted that, under simulation conditions similar to those considered in the present study, when the constraint in Equation~(\ref{within}) was not imposed and the variance at the initial time point ($\sigma_{w1}^2$) was uniquely estimated for the within-person process under the assumption of time-invariant residual variance—that is, when the number of free parameters increased to five—ML tended to become faster, whereas TS-MDFA tended to become slower. This difference may be attributable to the fact that ML no longer required estimation subject to this constraint, whereas, in TS-MDFA, the increase in the total number of parameters increased the computational burden because the parameters related to within-person variability (including $\sigma_{w1}^2$) must be updated iteratively by using ULS criterion. Thus, the computation time of TS-MDFA may be relatively sensitive to the model specifications. For further details, see the preprint corresponding to the initial submission of this paper.\footnote{https://arxiv.org/pdf/*****}}

\subsubsection{Bias, RMSE \textcolor{black}{and coverage} of the estimates}

\begin{figure}[htbp]
\textcolor{black}{
\centering
\includegraphics[width=\linewidth]{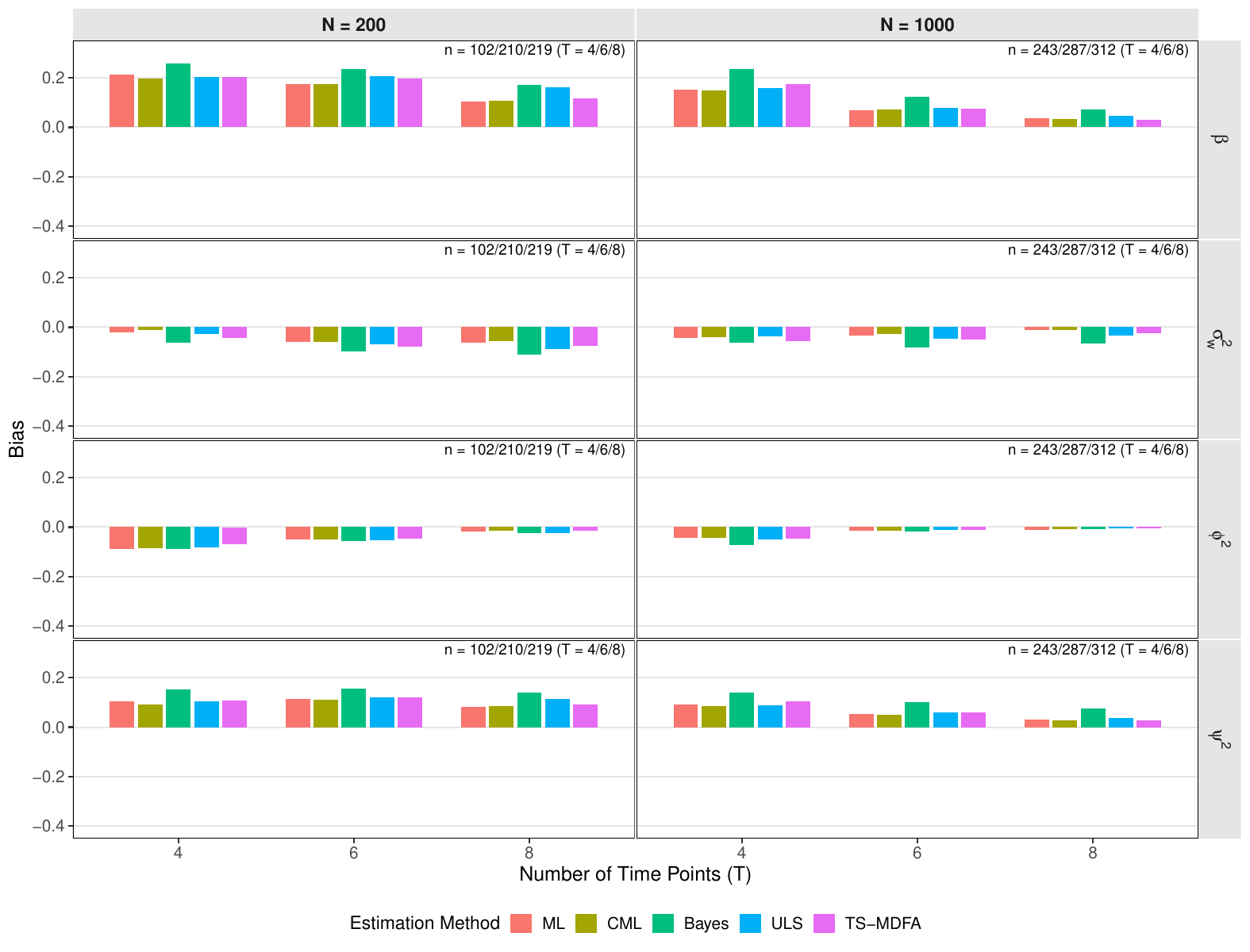}
\caption{Bias estimates by parameter type across methods ($\beta=0.3$, $\psi^2 = 0.1$). \textit{Note.} The number shown in the upper-right corner of each panel indicates the number of cases used to compute bias, i.e., the number of cases in which admissible solutions were observed for all methods.}
\label{fig:bias_by_param_psi1}
}
\end{figure}

\begin{figure}[htbp]
\textcolor{black}{
\centering
\includegraphics[width=\linewidth]{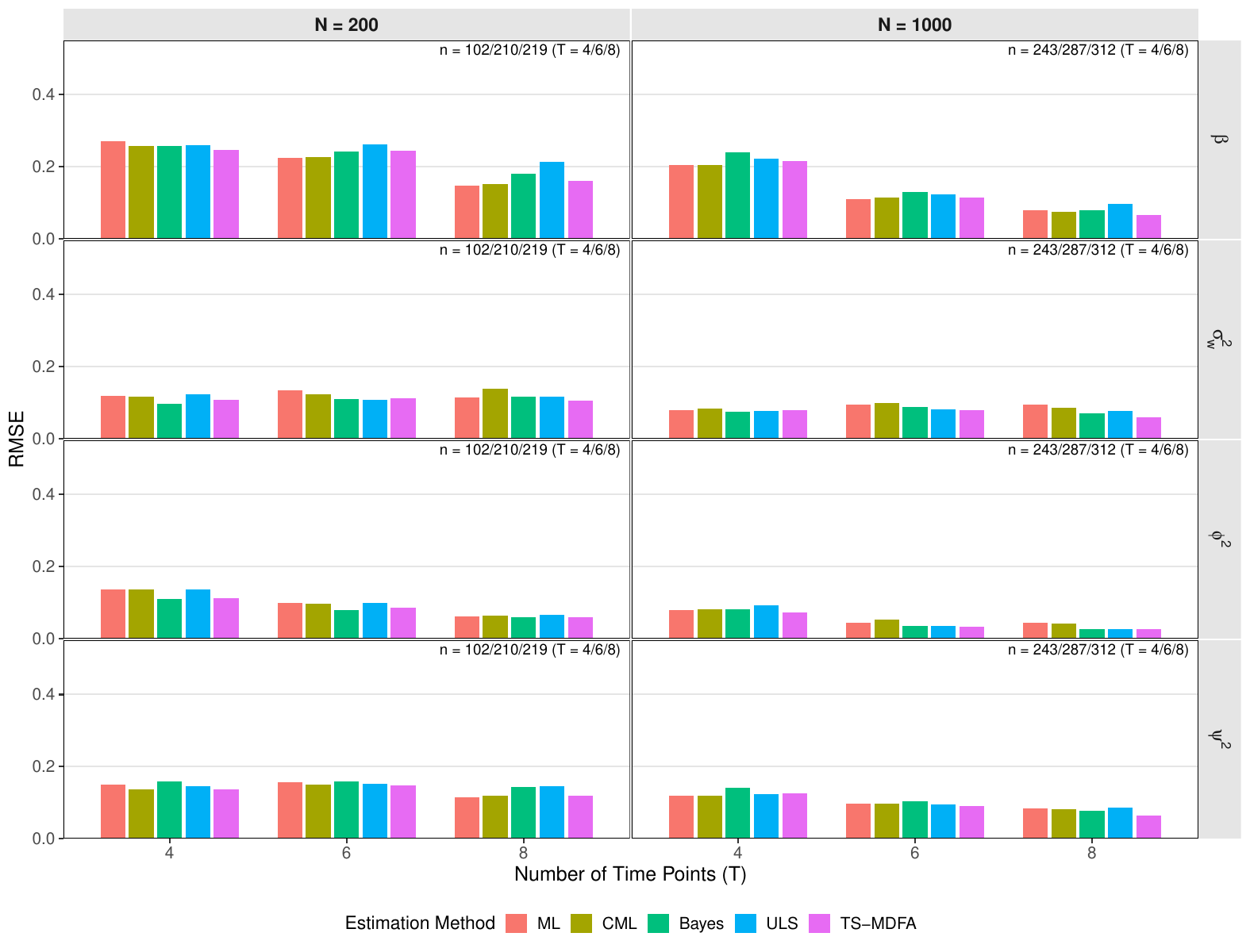}
\caption{RMSE estimates by parameter type across methods ($\beta=0.3$, $\psi^2 = 0.1$). \textit{Note.} The number shown in the upper-right corner of each panel indicates the number of cases used to compute RMSE, i.e., the number of cases in which admissible solutions were observed for all methods.}
\label{fig:RMSE_by_param_psi1}
}
\end{figure}

\begin{figure}[htbp]
\textcolor{black}{
\centering
\includegraphics[width=\linewidth]{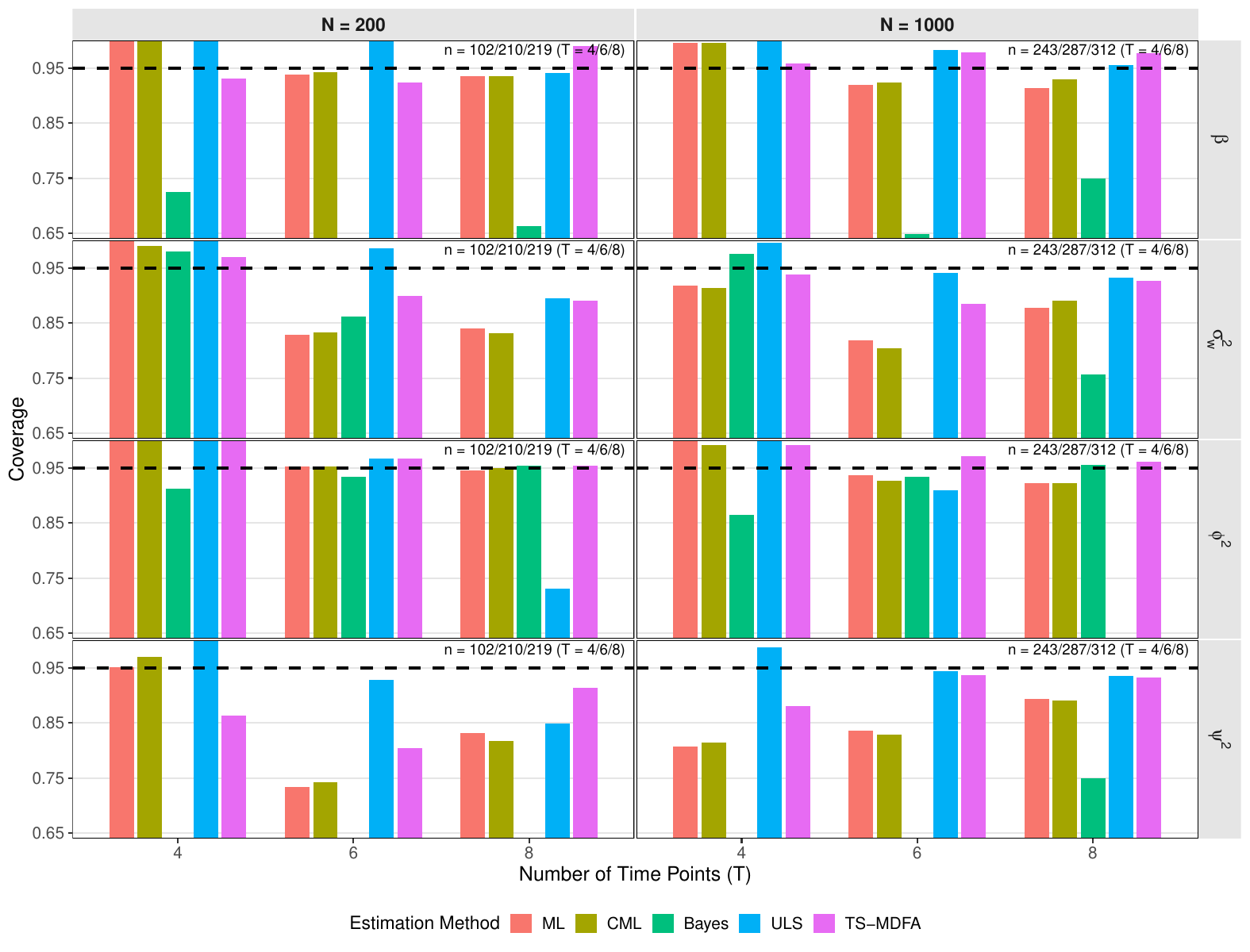}
\caption{Coverages by parameter type across methods ($\beta=0.3$, $\psi^2 = 0.1$). \textit{Note.} The number shown in the upper-right corner of each panel indicates the number of cases used to compute coverages, i.e., the number of cases in which admissible solutions were observed for all methods.}
\label{fig:coverage_by_param_psi1}
}
\end{figure}

Figures~2, ~3 \textcolor{black}{and 4} summarize the bias, \textcolor{black}{RMSE and coverage} of each estimation method across conditions by parameter type under the condition \textcolor{black}{$\beta=0.3$ and $\psi^2 = 0.1$}. The results were restricted to cases in which admissible solutions were observed for all methods. Specifically, in $N = 200$, the numbers of such cases were \textcolor{black}{102 for $T = 4$, 210 for $T = 6$, and 219 for $T = 8$, whereas in $N = 1{,}000$, these were 243 for $T = 4$, 287 for $T = 6$, and 312 for $T = 8$}.

As shown in Figure~2, bias estimates tended to decrease across all methods as both $N$ and $T$ increased. \textcolor{black}{Nestler et al. (2025) also showed similar tendencies when Bayesian estimation was used. Also,} under the condition $T = 8$ and $N = 1{,}000$, bias levels were pragmatically negligible in many methods. Among the methods, TS-MDFA exhibited \textcolor{black}{bias magnitudes generally comparable to those of the commonly used ML (and CML), while showing} the best overall performance. In contrast, the Bayesian estimation yielded relatively larger biases, indicating a non-negligible influence of the prior distributions. 
\textcolor{black}{Although the biases of all methods were generally smaller when $\beta = 0.5$, the patterns of differences among the methods described above remained largely similar. In particular, relative to the other estimation methods, Bayesian estimation exhibited pronounced bias when $\psi^2 = 0.1$ (see Figures~S1--S3 in the Online Supplemental Material).}
\textcolor{black}{The finding that relatively large biases were observed in the Bayesian estimates of parameters other than $\phi^2$ when $\beta$ was relatively small, and that these biases tended to be larger than those obtained with ML when $\phi^2$ was no larger than $\sigma_w^2$ (i.e., $\phi^2:\sigma_w^2 = 1:1$) as in the current setting, is consistent with L{\"u}dtke et al.\ (2018, Table 4).}

With regard to RMSE, Figure~3 indicates a general tendency for RMSE estimates to decrease as $N$ increased. On the other hand, for the measurement error variance parameter ($\psi^2$) \textcolor{black}{and the autoregressive coefficient parameter ($\beta$)}, the RMSEs remained non-negligible \textcolor{black}{relative to their true values} even when $T=8$ and $N=1{,}000$, highlighting the \textcolor{black}{relative difficulty of} obtaining stable estimates of these parameters.
\textcolor{black}{Again, TS-MDFA yielded RMSE values comparable to those of ML (and CML) across conditions, and the differences among the methods were not necessarily pronounced overall in Figure~3. However, Bayesian estimation yielded the smallest RMSEs overall when $\psi^2=0.3$, suggesting that the use of prior distributions helped mitigate the risk of extreme estimates. In addition, ML and CML tended to yield larger RMSEs when $\beta=0.5$ (see Figures~S4--S6 in the Online Supplemental Material). Overall, the RMSEs of TS-MDFA remained relatively small across conditions, indicating its favorable performance.}

\textcolor{black}{Figure~4 presents the coverage results for each estimation method, separately by parameter, when $\beta=0.3$ and $\psi^2=0.1$. Overall, TS-MDFA showed satisfactory coverage performance. However, for $\psi^2$, which was particularly susceptible to improper solutions, coverage tended to fall below the nominal level, especially when $N=200$. ML, CML, and ULS also generally showed satisfactory coverage, although their coverage tended to exceed the nominal level for many parameters when $T=4$. In contrast, Bayesian estimation yielded coverage rates substantially below the nominal level under many conditions, although this pattern was not consistently observed across all conditions; rather, a tendency toward overcoverage was observed in some conditions, particularly when $\psi^2 = 0.3$ (see Figures~S7--S9 in the Online Supplemental Material). This finding suggests that inferential results, including coverage rates, may be strongly affected by the choice of prior distributions. The observation that Bayesian estimation exhibited both overcoverage and undercoverage depending on the condition is consistent with L{\"u}dtke et al.\ (2018, Table~4). The tendency of ML, CML, and ULS to yield coverage above the nominal level when $T=4$ was also observed when $\beta=0.5$ or $\psi^2=0.3$. Taken together, TS-MDFA exhibited coverage closest to the nominal level overall.}

\textcolor{black}{The tendencies observed for the Bayesian estimation---namely, relatively larger biases and comparatively smaller RMSE than TS-MDFA---were confirmed even when these estimates were computed using all 500 replications in each condition. In contrast, when coverage was evaluated using all 500 replications, Bayesian estimation generally showed better performance than TS-MDFA, particularly when $N = 200$, although both methods tended to exhibit undercoverage. For TS-MDFA in particular, this may be partly attributable to the fact that the coverage rates were calculated while including replications that resulted in improper solutions. Nevertheless, the coverage performance of both methods improved as $T$ and $N$ increased (see Figures S10--S12 in the Online Supplemental Material for these results).}

\subsubsection{Correlations among estimates}
\begin{figure}[htbp]
\textcolor{black}{
\centering
\includegraphics[width=\linewidth]{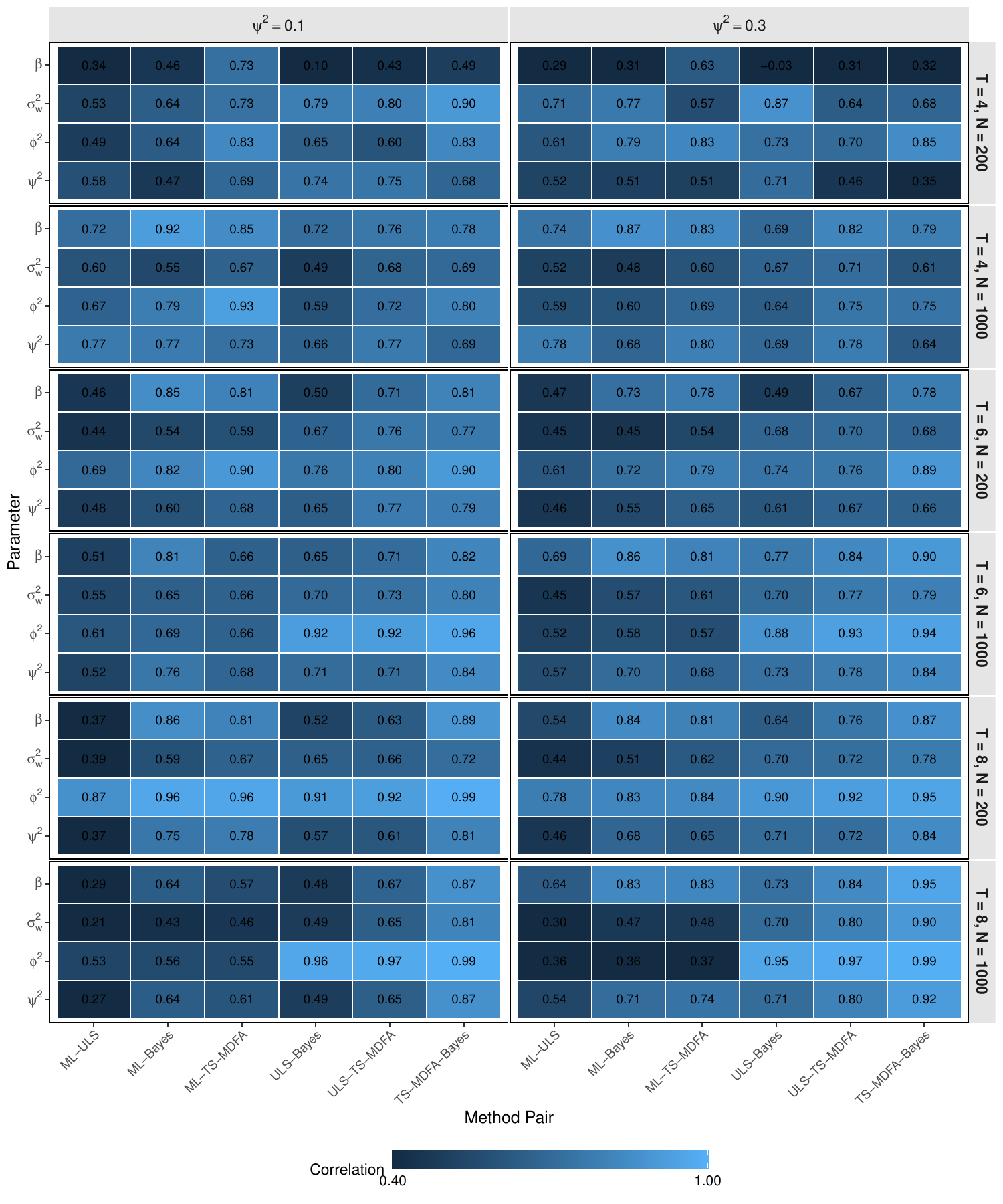}
\caption{Correlations of parameter estimates across method pairs ($\beta=0.3$).}
\label{fig:correlations}
}
\end{figure}

Figure~5 presents the correlations of parameter estimates across method pairs \textcolor{black}{when $\beta=0.3$}, restricted to cases in which admissible solutions were obtained for all methods. Because the correlations between ML and CML exceeded \textcolor{black}{0.90 in many parameters}, results for method pairs involving CML are omitted. 

Overall, \textcolor{black}{TS-MDFA exhibited higher correlations with Bayesian estimation and ULS, particularly as $T$ increased, with the correlations exceeding 0.80 or even 0.90 for each parameter. Compared with the correlations for these method pairs, ML tended to show lower correlations with Bayesian estimation and TS-MDFA on average, particularly when $N$ or $T$ was large. A similar pattern was observed when $\beta=0.5$ (see Figure~S13 in the Online Supplemental Material)}.

\textcolor{black}{This finding is not consistent with previous research, such as Adachi et al. (2019), who showed that MDFA produces estimates that are empirically very similar to those obtained using ML. We therefore examined the estimates from each replication more closely. Although these replications were not classified as yielding improper solutions, a small proportion of them (approximately 5\%) included variance parameter estimates from ML that were somewhat extremely underestimated or overestimated, presumably because of instability in the estimation. After excluding these replications, the correlations between ML and TS-MDFA exceeded 0.90 in each parameter. Thus, the observed discrepancies between these methods appear to be attributable primarily to the instability of the STARTS model estimation, particularly when using ML.}
\subsubsection{Summary and discussion}
In the estimation of the STARTS model using ML, CML and ULS, improper solutions occurred with a non-negligible frequency. In particular, improper solutions were highly frequent when both $T$ and $N$ were small, exceeding 50\%. In contrast, TS-MDFA and the Bayesian estimation were effective in avoiding improper solutions. However, TS-MDFA \textcolor{black}{shows a non-negligible tendency to produce relatively small variance estimate} (e.g., $\hat{\psi}^2 < .01$), \textcolor{black}{with the proportion of such cases reaching 10\%-20\% when $N$ was small}. Moreover, the correlations between estimates obtained from TS-MDFA and the Bayesian estimation were not necessarily high, in some cases falling below 0.90 \textcolor{black}{especially when $T$ and $N$ were small}.

From a computational perspective, \textcolor{black}{although its efficiency may be sensitive to the model specifications, TS-MDFA was the fastest method overall. In contrast, Bayesian estimation was the most computationally demanding, and its computation time increased substantially as $N$ increased. If sensitivity analyses using different prior distributions are conducted, the overall computational cost of Bayesian estimation becomes even greater.}

In terms of bias\textcolor{black}{, RMSE and coverage, TS-MDFA exhibited favorable performance, with values comparable to those of ML (and CML), whereas Bayesian estimation showed non-negligible sensitivity to the choice of prior distributions.}

Taken together, these results primarily indicate that TS-MDFA can serve as an effective method for the problem of improper solutions that frequently arise with ML, CML and ULS in estimating the STARTS model. Moreover, TS-MDFA may be preferable when bias arising from the choice of prior distributions is a concern or when it is difficult to specify appropriate (weakly informative) priors. In addition, from the perspective of sensitivity analysis, by referring to estimation results from TS-MDFA, researchers may be better able to delineate a plausible range of conclusions regarding the parameters of the STARTS model.
\section{Empirical Example}
\subsection{Data}
In this section, we present an empirical illustration of the TS-MDFA approach using data from the Tokyo Teen Cohort (TTC) study (Ando et~al., 2019), and we also compare the results with those obtained from the other estimation methods (i.e., ML, CML and ULS in SEM, as well as Bayesian 
estimation). The TTC study is a longitudinal cohort project designed to examine the psychological and physical development of adolescents ($N=3{,}171$) living in the Tokyo metropolitan region. Data were collected across $T=4$ measurement waves (ages 10, 12, 14 and 16) beginning in 2012, and a follow-up survey at age 20 has been conducted since 2022.

Here, we fit the STARTS model to the $T=4$ longitudinal data on sleep duration. Sleep duration in hours was assessed using the question ``How long do you usually sleep on weekdays?'' Because this item is based on self-reports covering a relatively broad period, it is reasonable to 
expect that a certain amount of measurement error influences on observations. 
In the current illustration, we analyze data from $N=1{,}294$ adolescents who provided complete responses to the sleep-duration questions across all four waves. The sample covariance and correlation matrix of sleep duration is provided in Table~\ref{tab:cov_cor_sd}.

\begin{table}[htbp]
\centering
\caption{Sample covariance and correlation matrix of sleep duration}
\label{tab:cov_cor_sd}
\begin{tabular}{lcccc}
\hline
        & SD10 & SD12 & SD14 & SD16 \\
\hline
SD10 & 0.394 & 0.546 & 0.342 & 0.167 \\
SD12 & 0.253 & 0.546 & 0.480 & 0.219 \\
SD14 & 0.185 & 0.304 & 0.738 & 0.430 \\
SD16 & 0.095 & 0.146 & 0.332 & 0.809 \\
\hline
\end{tabular}
\vspace{1mm}
\begin{flushleft}
\footnotesize
\textit{Note}. Diagonal elements and elements below the diagonal represent variances and covariances, respectively, whereas elements above the diagonal represent Pearson correlation coefficients.
\end{flushleft}
\end{table}
\subsection{Method}
As in the simulation study, when fitting the STARTS model, ML, CML and ULS estimations were conducted using the \texttt{lavaan} package (version~0.6-20; Rosseel, 2012), whereas Bayesian estimation was \textcolor{black}{performed using a custom-written \textsf{Stan} model implemented through the \texttt{rstan} package}.

In the TS-MDFA, we generated $M = 500$ sets of initial values and conducted estimation under each set. Initial values were sampled from $\phi^2 \sim \mathrm{gamma}(1,1), \psi^2 \sim \mathrm{gamma}(1,1), \beta \sim N(0.5,16),$ and $\sigma_{w}^2 \sim \mathrm{gamma}(1,1)$. Among the \textcolor{black}{resulting admissible solutions, we selected} the one that \textcolor{black}{yielded} the smallest value of the loss function $T_{\mathrm{MDFA}}(\bd{\theta})$ and \textcolor{black}{adopted} it as the final parameter estimate.

In the Bayesian estimation, prior distributions were specified as: $\phi^2 \sim \mathrm{Gamma}(1,1)$, $\psi^2 \sim \mathrm{Gamma}(1,1)$, and $\beta \sim N(0.5, 4^2)$, and $\sigma_w^2 \sim \mathrm{Gamma}(1,1)$. Posterior sampling was performed using HMC implemented in \textsf{Stan}. The warm-up phase was set to 5,000 iterations, followed by 50,000 sampling iterations \textcolor{black}{per chain}, using two independent Markov chains. The posterior means were used as the parameter estimates. From the perspective of the potential scale reduction statistic $\hat{R}$, we confirmed that there were no issues.

\subsection{Result}
\begin{table}[tb]
\textcolor{black}{
  \centering
  \caption{Point estimates and standard errors for parameters in the STARTS model obtained from each estimation method. Standard errors in the Bayesian estimation correspond to posterior standard deviations. The values shown in bold indicate improper solutions.}
  \label{tab:starts_estimates}
  \begin{tabular}{c*{4}{cc}cc}
    \hline
    & \multicolumn{2}{c}{ML} 
    & \multicolumn{2}{c}{CML}
    & \multicolumn{2}{c}{ULS}
    & \multicolumn{2}{c}{Bayes}
    & \multicolumn{2}{c}{TS-MDFA} \\
    \cline{2-11}
    Parameter 
      & Est. & SE 
      & Est. & SE 
      & Est. & SE 
      & Est. & SD 
      & Est. & SE\\
    \hline
    $\phi^2$ 
      & 0.147 & 0.017 
      & 0.026 & 0.058
      & 0.013 & 0.165 
      & 0.042 & 0.026
      & 0.074 & 0.019 \\
    $\psi^2$ 
      & \bd{-12.516} & 132.564 
      & 0.082 & 0.065    
      & 0.081 & 0.188 
      & 0.069 & 0.037    
      & 0.010 & 0.003 \\
    $\sigma_{w}^2$ 
      & 12.985 & 132.553
      & 0.511 & 0.038 
      & 0.529 & 0.077 
      & 0.510 & 0.035    
      & 0.536 & 0.030 \\
    $\beta$ 
      & 0.010 & 0.105 
      & 0.517 & 0.115 
      & 0.537 & 0.319 
      & 0.488 & 0.054 
      & 0.405 & 0.034 \\
    \hline
  \end{tabular}
}
\end{table}

Table~4 presents the point estimates and standard errors for parameters 
$\bd{\theta}$ obtained from each estimation method. Under the same criterion as in the simulation study (i.e., estimates smaller than 0.0001), improper solutions occurred in ML, but not in \textcolor{black}{CML}, ULS, Bayes or TS-MDFA.

In Bayes, the posterior standard deviations tended to be smaller than the standard errors obtained from \textcolor{black}{CML and} ULS, which may be attributable not only to the influence of the prior distributions.
\textcolor{black}{TS-MDFA also yielded relatively small standard errors for this dataset, although the pattern of the estimates differed somewhat from that of the Bayesian estimation.}

\textcolor{black}{CML, ULS, and the Bayesian estimation yielded similar parameter estimates. In contrast, TS-MDFA produced a relatively larger estimate of $\phi^2$, whereas the estimates of $\psi^2$ and $\beta$ were smaller than those obtained by the other methods. In contrast, the estimates of $\sigma_w^2$ were similar across methods, suggesting that a substantial proportion of the variance in the outcomes was attributable to within-person variability at each time point.}

As noted above, given the self-report nature of the measurements and the influence of day-to-day fluctuations in sleep duration, it is conceptually reasonable to assume that the observations contain non-zero measurement error. \textcolor{black}{From this perspective, in TS-MDFA estimate of the measurement error variance ($\hat{\psi}^2 = 0.010$) appears to be rather small}. Conversely, given that stable individual differences in sleep duration during adolescence are well documented and are known to depend on factors such as constitution and living environment, \textcolor{black}{the estimates of the stable trait factor variance ($\phi^2$) obtained by CML, ULS, and the Bayesian estimation also appear to be implausibly small. The estimates of $\beta$ ranged approximately from 0.40 to 0.55 across methods}, indicating that within-person deviations in sleep duration at a given occasion retained a moderate degree of temporal persistence until the subsequent measurement occasion (two years later).

To further examine the characteristics of the estimates obtained by each method from the perspective of local fit (e.g., Kline, 2023), Table~5 presents the Bentler-type residual correlation matrices, in which both the observed and model-implied covariance matrices were rescaled by dividing each element by the square root of the product of the corresponding observed variances, computed from the parameter estimates obtained by each method. \textcolor{black}{Interestingly, although noticeable differences in the point estimates of several parameters were observed across methods (Table~4; between TS-MDFA and other methods), the model-implied covariance matrices, and consequently the residual correlation matrices, were remarkably similar across estimation methods. Nevertheless, all methods consistently exhibited relatively large residuals for the variance at the first measurement occasion. As shown in Table~3, the variance of sleep duration increased over time, whereas the STARTS model assumes time-invariant variance parameters and, consequently, time-invariant outcome variances. This mismatch between the model assumptions and the data appears to account for the poor fit. Consistent with this interpretation, all methods yielded SRMR values of approximately 0.18, indicating unsatisfactory global model fit.}

\textcolor{black}{To address this issue, we relaxed the standard assumptions of the STARTS model. Specifically, following a strategy commonly adopted in the RI-CLPM literature, we allowed the within-person variance to vary over time by removing the constraint imposed by Equation~(\ref{within}). In addition to the (time-invariant) residual variance $\omega^2$, the within-person variance at the first measurement occasion ($\sigma_{w1}^2$) was introduced as an additional free parameter. Under this modified model, CML resulted in an improper solution in $\hat{\psi^2}=0.000$, whereas ULS, the Bayesian estimation, and TS-MDFA yielded admissible solutions and showed substantial improvements in both global and local fit, with SRMR values ranging from approximately 0.07 to 0.09. Notably, TS-MDFA no longer produced extremely small estimates of either $\hat{\psi}^2$ or $\hat{\phi}^2$, resulting in parameter estimates that appeared considerably more interpretable. Moreover, although TS-MDFA yielded point estimates similar to those obtained using Bayesian estimation, its standard errors were larger in TS-MDFA. Further details of the estimation results are provided in the Table~S5 of the Online Supplemental Material.}

As illustrated in this example, TS-MDFA can flexibly estimate the parameters of the STARTS model even in situations where commonly used methods such as ML produce improper solutions. 
Moreover, because TS-MDFA allows estimation without requiring the specification of prior distributions while still yielding plausible solutions, it can be used not only as a standalone estimation approach but also as a tool for sensitivity analysis that compares estimation results obtained from different methods.

\begin{table}[htbp]
\textcolor{black}{
\centering
\caption{Residual correlation matrices obtained under \textcolor{black}{CML,} ULS, Bayesian estimation and TS-MDFA. In each element, the residual correlations are provided in the order \textcolor{black}{CML/}ULS/Bayesian estimation/TS-MDFA.}
\label{tab:residual_correlations}
\setlength{\tabcolsep}{3pt}
\hspace*{-1.5cm}
\begin{tabular}{lcccc}
\hline  
      & SD10 & SD12 & SD14 & SD16 \\
\hline
SD10  & -0.571/-0.577/-0.575/-0.574      &       &       &       \\
SD12  & -0.078/-0.093/-0.081/-0.082   & -0.135/-0.139/-0.136/-0.137  &       &       \\
SD14  & 0.042/0.036/0.039/0.041  & 0.023/0.012/0.022/0.020 & 0.160/0.157/0.160/0.159 &       \\
SD16  & -0.003/0.000/-0.012/-0.027 & -0.024/-0.029/-0.027/-0.025 & 0.055/0.046/0.054/0.052  & 0.235/0.232/0.234/0.233 \\
\hline 
\end{tabular}
}
\end{table}

\section{General Discussion}
In the present study, motivated by the idea of reformulating the STARTS model as a factor-analytic representation in order to apply the MDFA framework, we proposed a new two-stage estimation approach (TS-MDFA). This approach addresses the problem of improper solutions that can arise when estimating the STARTS model via eigen-decomposition, and it may also be effective in mitigating possible estimation bias associated with Bayesian estimation. \textcolor{black}{In addition, whereas TS-MDFA was the fastest method overall in the present simulation, Bayesian estimation can be computationally demanding.}

TS-MDFA can also be interpreted as an extension of MDFA to SEMs that encompass certain submodels, such as the STARTS model. In contrast to approaches that rely on regularization, such as the one proposed by Yamashita (2024), TS-MDFA provides a more natural inferential framework without incorporating additional penalty terms.

TS-MDFA adopts a two-stage estimation procedure in which the model parameters are partitioned into two subsets, $\bd{\theta_1}$ and $\bd{\theta_2}$, and updated separately. More specifically, the cross-covariance matrix $\bd{S_{Y\tilde{Z}}}$, which reflects the covariance structure between the observed data $\bd{Y}$ and the (exogenous) common and unique factors $\bd{\tilde{Z}}$ in the STARTS model, is updated using eigen-decomposition (Step 2). Based on the updated $\bd{S_{Y\tilde{Z}}}$, the subvector $\bd{\theta_1}$, which is associated with within-person variability, is then updated under the ULS criterion (Step 3.1). Subsequently, the remaining subvector $\bd{\theta_2}$ is updated in a manner analogous to the \textcolor{black}{original} MDFA procedure (Step 3.2). Under the MDFA loss function, these steps are iterated, and the optimal solution is chosen using a multiple-start strategy to mitigate the possible risk of convergence to local minima.

In the simulation study and illustrative example, we demonstrate that the TS-MDFA substantially reduces the occurrence of improper solutions compared with ML and ULS, without requiring the specification of prior distributions. The method can be used not only as a standalone estimation approach but also as a tool for sensitivity analysis that compares estimation results obtained from different methods.

As described above, TS-MDFA can be viewed as an extension of MDFA-based estimation to SEMs that explicitly includes structural equations among observed and latent variables. Consequently, TS-MDFA is applicable not only to a wide range of submodels that can be described within the SEM framework, but also to extended methods built upon those models. As one example, latent growth models (LGMs; Meredith \& Tisak, 1990) are empirically known to be prone to improper solutions, particularly when $T$ is small or when the functional form of the trajectory (e.g., linear or quadratic) is misspecified (e.g., Usami, 2026). This problem can be addressed by a new estimation approach like TS-MDFA. Improper solutions in LGMs become an especially serious issue in contexts where model estimation must be repeated a large number of times, such as in SEMTree approaches (e.g., Brandmaier et al., 2013), which aim to identify heterogeneous subpopulations characterized by different parameter values and covariates that explain such heterogeneity. The application of TS-MDFA to SEMTree constitutes an interesting direction for future research, and our research group is currently engaged in the development and empirical validation of such extensions.

In addition, although mathematically simpler than the STARTS model, RI-CLPM, which has seen explosive growth over the past decade in applications particularly for the purpose of inferring reciprocal relations between variables, also tend to suffer from improper solutions. This risk is expected to increase as the number of variables increases, and especially in such cases the application of TS-MDFA is highly promising for \textcolor{black}{estimating} the RI-CLPM as well. 

In this sense, TS-MDFA offers a new direction for estimation in both the STARTS model and the RI-CLPM, and it is expected to be extendable to and applicable across a wide variety of SEM-based analyses. At the same time, 
several important issues and potential limitations remain unexamined in the present study and should be addressed in future work. One such issue concerns the asymptotic properties of the proposed two-stage estimation with respect to $T$ and $N$. Under the original MDFA framework, the strong consistency of the estimator has been established (Terada, 2025; Theorem~3.4). It is required to provide a mathematical proof of consistency in TS-MDFA, taking into account comparisons with other estimation methods.

Moreover, although TS-MDFA evaluates candidate solutions based on the loss function given in Equation~(\ref{MDFAloss}) together with a predetermined patience criterion, alternative evaluation strategies may also be considered. In particular, because TS-MDFA combines the MDFA-based updating for $\bd{\theta_2}$ with the SEM-based ULS updating for $\bd{\theta_1}$, it is conceivable to evaluate solutions using alternative loss functions, such as ULS in SEM (Equation~(\ref{ULS})). In this case, the resulting estimates are expected to be more similar to those obtained via ULS estimation. Relatedly, compared with the standard ML estimation, TS-MDFA involves several hyperparameters, such as the numbers of patience and multiple starts. Although the required numbers of these hyperparameters are expected to decrease as $T$ and $N$ increase, more principled guidelines for their selection remain an open topic for future investigation. Furthermore, although missing data were not considered in the present study, computational efficiency may become a critical issue if TS-MDFA is combined with multiple imputation procedures to account for missing data assumed to be MAR. \textcolor{black}{It may also be possible to develop an approach that, analogous to direct likelihood estimation under ML, estimates the model parameters based on the sample covariance matrices obtained from subgroups of individuals sharing the same missing data pattern.} Further research aimed \textcolor{black}{at evaluating the performance of these approaches and} at improving algorithmic efficiency should therefore be important.

Next, although extending TS-MDFA to cases involving two or more variables, such as the RI-CLPM, is technically straightforward, the risk of improper solutions is expected to increase due to potential instability in the estimated covariance matrices associated with stable trait factors (e.g., Usami, Todo et al., 2019). Accordingly, it will be necessary to examine the usefulness of TS-MDFA in multivariate settings.

Both MDFA and TS-MDFA assume that the data $\bd{Y}$ are continuous. However, in many applications of the STARTS models as well as LGMs, ordinal categorical data are frequently encountered. In such cases, SEM commonly employs a two-stage estimation approach, in which polychoric correlations among variables are first computed and the model parameters are then estimated using (diagonal) WLS criterion based on these correlations (e.g., Rhemtulla et al., 2012). From the perspective of extending (TS-)MDFA to non-metric (TS-)MDFA, it would be a highly interesting research topic to investigate how the frequency of improper solutions and the properties of the resulting estimates differ between such extensions and the above traditional approaches.

Last but not least, due to the computational cost associated with certain conditions---particularly those involving Bayesian estimation---the conditions examined in the simulations conducted in this study are still limited. Accordingly, further simulation studies are required to examine performance of methods in a broader range of conditions.

SEM-based approaches to longitudinal data analysis provide an exceptionally attractive framework that encompasses a wide range of analytical models and empirically testable research hypotheses. At the same time, the potential problem of improper solutions has remained a major obstacle. It is our hope that the proposed TS-MDFA will help mitigate this issue, thereby increasing awareness of the appeal of longitudinal SEM among researchers and practitioners and further expanding its range of applications in the future.

\newpage
\hspace{-6mm}{\bf{{\Large{References}}}}\\
Adachi, K., Ito, M., Uno, K. (2019). Matrix decomposition factor analysis and its new\par
developments. \textit{Bulletin of the Computational Statistics of Japan, 32}(1), 61-77.\par
\hspace{-6mm}Adachi, K., \& Trendafilov, N.T. (2018). Some mathematical properties of the matrix\par
decomposition solution in factor analysis. \textit{Psychometrika, 83}, 407-424.\par
\hspace{-6mm}Andersen, H.K. (2022). Equivalent approaches to dealing with unobserved heterogeneity\par
in cross-lagged panel models: Investigating the benefits and drawbacks of the latent curve\par
model with structured residuals and the random intercept cross-lagged panel model.\par
\textit{Psychological Methods, 27}(5), 730-751.\par
\hspace{-6mm}Ando, S. et al. (2019). Cohort profile: Tokyo Teen Cohort study (TTC). \textit{International}\par
 \textit{Journal of Epidemiology, 48}, 1414-1414g.\par
\hspace{-6mm}Brandmaier, A.M., Oertzen, T.V., McArdle, J.J., \& Lindenberger, U. (2013). Structural\par equation model trees. \textit{Psychological Methods, 18}, 71-86.\par
\hspace{-6mm}\textcolor{black}{Browne, M.W. (1984). Asymptotically distribution-free methods for the analysis of covariance}\par
\textcolor{black}{structures. \textit{British Journal of Mathematical and Statistical Psychology, 37}(1), 62-83.}\par
\hspace{-6mm}Dhaene, S. \& Rosseel, Y. (2023). An evaluation of non-iterative estimators in the structural\par
after measurement (SAM) approach to structural equation modeling (SEM).\par \textit{Structural Equation Modeling: A Multidisciplinary Journal, 30}(6), 926-940.\par
\hspace{-6mm}Dicke, T., Parker, P.D., Guo, J., Basarkod, G., Marsh, H.W., Deady, M., Harvey, S., \& Riley,\par
P. (2022). Ubiquitous emotional exhaustion in school principals: Stable trait, enduring\par autoregressive trend, or occasion-specific state? \textit{Journal of Educational Psychology, 114}(2),\par
426–441. https://doi.org/10.1037/edu0000582\par
\hspace{-6mm}\textcolor{black}{Dominitz, J., \& Sherman, R.P. (2005). Some convergence theory for iterative estimation procedures}\par
\textcolor{black}{with an application to semiparametric estimation. \textit{Econometric Theory, 21}(4), 838-863.}\par
\hspace{-6mm}Du, H., \& Bentler, P.M. (2022). 40-year old unbiased distribution free estimator reliably\par improves SEM statistics for nonnormal data. \textit{Structural Equation Modeling: }\par \textit{A Multidisciplinary Journal, 29}(6), 872-887.\par
\hspace{-6mm}Hamaker, E.L. (2023). The within-between dispute in cross-lagged panel research and how\par
to move forward. \textit{Psychological Methods}. Advance online publication.\par
https://doi.org/10.1037/met0000600\par
\hspace{-6mm}Hamaker, E.L., Kuiper, R.M., \& Grasman, R.P.P.P. (2015). A critique of the cross-lagged\par
panel model. \textit{Psychological Methods, 20}(1), 102-116.\par
\hspace{-6mm}Hayakawa, K., \& Sun, Q. (2022). Selection of loss function in covariance structure analysis:\par 
Case of the spherical model. \textit{Structural Equation Modeling: A Multidisciplinary Journal,}\par 
\textit{29}(4), 507-520.\par
\hspace{-6mm}Kaplan, D. (2009). \textit{Structural equation modeling: Foundations and extensions (2nd ed.)}. Sage.\par
\hspace{-6mm}Kenny, D.A., \& Zautra, A. (1995). The trait-state-error model for multiwave data.\par
\textit{Journal of Consulting and Clinical Psychology, 63}(1), 52-59.\par
\hspace{-6mm}Kenny, D.A., \& Zautra, A. (2001). Trait–state models for longitudinal data. In L.M. Collins \& \par A.G. Sayer (Eds.), New methods for the analysis of change (pp. 243–263). Washington, DC:\par
American Psychological Association. http://dx.doi.org/10.1037/10409-008\par
\hspace{-6mm}Kline, R.B. (2023). \textit{Principles and practice of structural equation modeling (5th ed.)}.\par 
Guilford Press.\par
\hspace{-6mm}Loh, W.W., \& Ren, D. (2023). A tutorial on causal inference in longitudinal data with\par
time-varying confounding using G-estimation. \textit{Advances in Methods and Practices in}\par
{\textit{Psychological Science, 6}(3), https://doi.org/10.1177/25152459231174029}\par
\hspace{-6mm}Loh, W.W., \& Ren, D. (2025). Estimating time-varying treatment effects in longitudinal\par
studies. \textit{Psychological Methods, 30}(2), 240-253.\par
\hspace{-6mm}Lucas, R.E. (2023). Why the cross-lagged panel model is almost never the right choice.\par
\textit{Advances in Methods and Practices in Psychological Science, 6}, 1-22.\par
\hspace{-6mm}L$\ddot{\mathrm{u}}$dtke, O., \& Robitzsch, A. (2022). A comparison of different approaches for estimating\par
cross-lagged effects from a causal inference perspective. \textit{Structural Equation Modeling:}\par \textit{A Multidisciplinary Journal,} \textit{29}(6), 888-907.\par
\hspace{-6mm}L$\ddot{\mathrm{u}}$dtke, O., \& Robitzsch, A. (2025, June 25). Against the ubiquity of the random intercept\par cross-lagged panel model. https://doi.org/10.31234/osf.io/ua8gnv1\par
\hspace{-6mm}L$\ddot{\mathrm{u}}$dtke, O., Robitzsch, A., \& Wagner, J. (2018). More stable estimation of the STARTS model:\par A Bayesian approach using Markov chain Monte Carlo techniques. \textit{Psychological Methods,}\par \textit{23}(3), 570-593.\par
\hspace{-6mm}McNeish, D., \& Hamaker, E.L. (2020). A primer on two-level dynamic structural equation\par
models for intensive longitudinal data in Mplus. \textit{Psychological Methods, 25}(5), 610-635.\par
\hspace{-6mm}Meredith, W., \& Tisak, J. (1990). Latent curve analysis. \textit{Psychometrika, 55}, 107-122.\par
\hspace{-6mm}Merkle, E.C., \& Rosseel, Y. (2018). blavaan: Bayesian structural equation models via\par parameter expansion. \textit{Journal of Statistical Software, 85}(4), 1-30.\par
\hspace{-6mm}Miočević, M., Levy, R., \& MacKinnon, D.P. (2021). Different roles of prior distributions\par in the single mediator model with latent variables. \textit{Multivariate Behavioral Research, 56}(1),\par
20-40.\par
\hspace{-6mm}\textcolor{black}{Nestler, S., \& Humberg, S. (2024). Univariate autoregressive structural equation models as}\par 
\textcolor{black}{mixed-effects models. \textit{Structural Equation Modeling: A Multidisciplinary Journal, 31}, 357-366.}\par
\hspace{-6mm}\textcolor{black}{Nestler, S., Robitzsch, A., \& Lüdtke, O. (2025). Fitting single- and multiple-indicator STARTS}\par
\textcolor{black}{models as Dynamic Structural Equation Models. \textit{Structural Equation Modeling: A}}\par
\textcolor{black}{ \textit{Multidisciplinary Journal, 32}, 529-540.}\par
\hspace{-6mm}Orth, U., Clark, D.A., Donnellan, M.B., \& Robins, R.W. (2021). Testing prospective effects in\par
longitudinal research: Comparing seven competing cross-lagged models. \textit{Journal of}\par \textit{Personality and Social Psychology, 120}(4), 1013-1034.\par
\hspace{-6mm}R Core Team. (2025). R: A Language and Environment for Statistical Computing. R Foundation\par for Statistical Computing.
https://www.r-project.org/\par
\hspace{-6mm}Rhemtulla, M., Brosseau-Liard, P.E., \& Savalei, V. (2012). When can categorical variables\par 
be treated as continuous? A comparison of robust continuous and categorical SEM estimation\par 
methods under suboptimal conditions. \textit{Psychological Methods, 17}(3), 354-373.\par
\hspace{-6mm}Rossell, Y. (2012). Lavaan: An R package for structural equation modeling. \textit{Journal of}\par
\textit{Statistical Software, 48}(2), 1-36.\par
\hspace{-6mm}Rosseel, Y., \& Loh, W.W. (2024). A structural after measurement approach to structural\par equation modeling. \textit{Psychological Methods, 29}(3), 561-588.\par
\hspace{-6mm}Searle, S.R., Casella, G., \& McCulloch, C.E. (1992). \textit{Variance components}. New York,\par
NY: Wiley. http://dx.doi.org/10.1002/9780470316856\par
\hspace{-6mm}Siepe, B.S., Bartoš, F., Morris, T.P., Boulesteix, A.-L., Heck, D.W., \& Pawel, S. (2024).\par Simulation studies for methodological research in psychology: A standardized template for\par 
planning, preregistration, and reporting. \textit{Psychological Methods}. Advance online\par
publication. https://doi.org/10.1037/met0000695\par
\hspace{-6mm}Smid, S.C., McNeish, D., Miočević, M., \& Schoot, R.V.D. (2020). Bayesian versus frequentist\par
estimation for structural equation models in small sample contexts: A systematic review.\par \textit{Structural Equation Modeling: A Multidisciplinary Journal, 27}(1), 131-161.\par
\hspace{-6mm}Stan Development Team. (2024). RStan: The R interface to Stan.\par
\hspace{-6mm}\textcolor{black}{Stefanski, L.A., \& Boos, D.D. (2002). The calculus of M-estimation. \textit{The American Statistician,}}\par
\textcolor{black}{\textit{56}(1), 29-38.}\par
\hspace{-6mm}Terada, Y. (2025). Statistical properties of matrix decomposition factor analysis.\par arXiv:https://arxiv.org/abs/2403.06968\\
\hspace{-6mm}Ulitzsch, E., L$\ddot{\mathrm{u}}$dtke, O., \& Robitzsch, A. (2023). Alleviating estimation problems in small sample\par structural equation modeling- A comparison of constrained maximum
likelihood, Bayesian\par estimation, and fixed reliability approaches. \textit{Psychological Methods, 28}, 527-557.\par
\hspace{-6mm}Usami, S. (2021). On the differences between general cross-lagged panel model and\par
random-intercept cross-lagged panel model: Interpretation of cross-lagged parameters\par
and model choice. \textit{Structural Equation Modeling: A Multidisciplinary Journal, 28}\par
(3), 331-344.\par
\hspace{-6mm}Usami, S. (2022). Statistical models for the inference of within-person relations:\par
A random intercept cross-lagged panel model and its interpretation. \textit{The Japanese}\par
\textit{Journal of Developmental Psychology, 33,} 267-286. (In Japanese)\par
\hspace{-6mm}Usami, S. (2023). Within-person variability score-based causal inference: A two-step\par
estimation for joint effects of time-varying treatments. \textit{Psychometrika, 88}(4),\par
1466-1494.\par
\hspace{-6mm}Usami, S. (2026). \textit{Longitudinal data analysis: A structural equation modeling approach.}\par
University of Tokyo Press (In Japanese).\par 
\hspace{-6mm}Usami, S., Murayama, K., \& Hamaker, E.L. (2019). A unified framework of longitudinal\par
models to examine reciprocal relations. \textit{Psychological Methods, 24}(5), 637-657.\par
\hspace{-6mm}Usami, S., Todo, N., \& Murayama, K. (2019). Modeling reciprocal effects in medical\par
research: Critical discussion on the current practices and potential alternative models.\par
\textit{PLOS ONE, 14(9)}: e0209133.\par
\hspace{-6mm}Wickham, H. (2016). ggplot2: Elegant graphics for data analysis. Springer.\par https://ggplot2.tidyverse.org\par
\hspace{-6mm}Yamashita, N. (2024). Matrix decomposition approach for structural equation modeling as an\par
alternative to covariance structure analysis and its theoretical properties.\par
\textit{Structural Equation Modeling: A Multidisciplinary Journal, 31}(5), 817-834.\par
\hspace{-6mm}Zheng, B.Q., \& Bentler, P.M. (2021). Testing mean and covariance structures with reweighted\par  least squares. \textit{Structural Equation Modeling: A Multidisciplinary Journal, 29}(2), 259-266.
\newpage

\clearpage
\thispagestyle{empty}

\vspace*{\fill}
\begin{center}
{\LARGE \bfseries Online Supplemental Material}
\end{center}
\vspace*{\fill}
\newpage
\clearpage
\thispagestyle{empty}
\setcounter{table}{0}
\setcounter{figure}{0}
\renewcommand{\thetable}{S\arabic{table}}

\begin{table}[htbp]
\textcolor{black}{
\centering
\caption{Proportions of improper solutions across estimation methods ($\beta = 0.5$).}
\label{tab:improper_beta_0p5}
\begin{tabular}{rrrrrrrrrr}
\toprule
$T$ & $N$ & $\psi^2$ & ML & ULS & CML & ALL & TS-MDFA & BAY & TS-MDFA($10^{-2}$) \\
\midrule
4 & 200 & 0.1 & 0.52 & 0.65 & 0.48 & 0.40 & 0.02 & 0.00 & 0.11 \\
4 & 200 & 0.3 & 0.69 & 0.75 & 0.60 & 0.51 & 0.01 & 0.00 & 0.14 \\
4 & 1000 & 0.1 & 0.17 & 0.27 & 0.15 & 0.12 & 0.00 & 0.00 & 0.07 \\
4 & 1000 & 0.3 & 0.27 & 0.34 & 0.25 & 0.20 & 0.00 & 0.00 & 0.01 \\
6 & 200 & 0.1 & 0.25 & 0.35 & 0.20 & 0.14 & 0.09 & 0.00 & 0.18 \\
6 & 200 & 0.3 & 0.19 & 0.28 & 0.17 & 0.11 & 0.01 & 0.00 & 0.13 \\
6 & 1000 & 0.1 & 0.07 & 0.10 & 0.05 & 0.02 & 0.00 & 0.00 & 0.06 \\
6 & 1000 & 0.3 & 0.02 & 0.02 & 0.02 & 0.00 & 0.00 & 0.00 & 0.00 \\
8 & 200 & 0.1 & 0.13 & 0.27 & 0.11 & 0.06 & 0.09 & 0.00 & 0.16 \\
8 & 200 & 0.3 & 0.09 & 0.15 & 0.07 & 0.04 & 0.01 & 0.00 & 0.06 \\
8 & 1000 & 0.1 & 0.03 & 0.04 & 0.03 & 0.00 & 0.00 & 0.00 & 0.01 \\
8 & 1000 & 0.3 & 0.02 & 0.01 & 0.02 & 0.00 & 0.00 & 0.00 & 0.00 \\
\bottomrule
\end{tabular}
\begin{flushleft}
\footnotesize Note. Improper solutions are defined as cases in which at least one variance-parameter estimate is less than 0.0001. ALL indicates the proportion of replications in which ML, ULS, and CML all yielded improper solutions. TS-MDFA($10^{-2}$) indicates the corresponding proportion for TS-MDFA when the threshold is changed to 0.01.
\end{flushleft}
}
\end{table}

\begin{table}[htbp]
\textcolor{black}{
\centering
\caption{Required computation time (in seconds) across estimation methods ($\beta = 0.5$).}
\label{tab:computation_time_beta_0p5}
\begin{tabular}{ccc|rrrrrrrrrr}
\hline
\multicolumn{3}{c|}{} &
\multicolumn{2}{c}{ML} &
\multicolumn{2}{c}{CML} &
\multicolumn{2}{c}{Bayes} &
\multicolumn{2}{c}{ULS} &
\multicolumn{2}{c}{TS-MDFA} \\
\hline
$T$ & $N$ & $\psi^2$ &
$M$ & $SD$ &
$M$ & $SD$ &
$M$ & $SD$ &
$M$ & $SD$ &
$M$ & $SD$ \\
\hline
4 & 200 & 0.1 & 12.5 & 12.9 & 12.1 & 15.0 & 19.7 & 1.1 & 17.4 & 18.0 & 5.7 & 5.8 \\
 &  & 0.3 & 16.0 & 15.7 & 14.2 & 16.5 & 20.5 & 1.4 & 16.6 & 15.5 & 7.3 & 7.0 \\
 & 1000 & 0.1 & 5.6 & 7.4 & 6.8 & 9.3 & 50.8 & 2.8 & 4.9 & 7.0 & 3.5 & 3.0 \\
 &  & 0.3 & 7.2 & 8.6 & 7.8 & 10.0 & 56.8 & 6.2 & 5.8 & 7.0 & 4.5 & 3.5 \\
\hline
6 & 200 & 0.1 & 21.6 & 28.1 & 27.1 & 33.7 & 21.1 & 1.1 & 20.0 & 21.7 & 5.8 & 5.5 \\
 &  & 0.3 & 24.4 & 28.2 & 25.2 & 30.2 & 22.6 & 1.7 & 19.3 & 18.9 & 6.3 & 6.0 \\
 & 1000 & 0.1 & 12.6 & 16.5 & 15.5 & 19.7 & 59.8 & 3.5 & 7.1 & 5.5 & 3.5 & 2.9 \\
 &  & 0.3 & 14.5 & 18.9 & 15.9 & 20.3 & 68.9 & 9.5 & 7.0 & 6.6 & 3.8 & 2.8 \\
\hline
8 & 200 & 0.1 & 32.3 & 40.1 & 40.4 & 50.5 & 22.3 & 1.8 & 26.5 & 23.3 & 4.6 & 4.5 \\
 &  & 0.3 & 34.4 & 43.2 & 39.0 & 46.7 & 25.2 & 1.9 & 29.8 & 26.2 & 5.2 & 4.8 \\
 & 1000 & 0.1 & 32.8 & 41.4 & 37.0 & 45.9 & 93.7 & 11.5 & 21.6 & 16.3 & 5.3 & 3.9 \\
 &  & 0.3 & 22.0 & 28.1 & 23.5 & 29.8 & 75.6 & 10.8 & 13.7 & 9.4 & 3.3 & 1.8 \\
\hline
\end{tabular}
\begin{flushleft}
\footnotesize
\textit{Note.} Result for TS-MDFA is based on 10 multiple starts. Results for Bayes are based on 11,000 sampling iterations for each of two independent Markov chains. $M$ = mean; $SD$ = standard deviation.
\end{flushleft}
}
\end{table}

\begin{table}[htbp]
\textcolor{black}{
\centering
\caption{Required computation time (in seconds) across estimation methods ($\beta = 0.3$; restricted to cases in which no improper solutions occurred for each method).}
\label{tab:computation_time_proper_only_beta_0p3}
\begin{tabular}{ccc|rrrrrrrrrr|r}
\hline
\multicolumn{3}{c|}{} &
\multicolumn{2}{c}{ML} &
\multicolumn{2}{c}{CML} &
\multicolumn{2}{c}{Bayes} &
\multicolumn{2}{c}{ULS} &
\multicolumn{2}{c|}{TS-MDFA} &
$n$ \\
\hline
$T$ & $N$ & $\psi^2$ &
$M$ & $SD$ &
$M$ & $SD$ &
$M$ & $SD$ &
$M$ & $SD$ &
$M$ & $SD$ &
\\
\hline
4 & 200 & 0.1 & 7.5 & 11.1 & 8.0 & 9.8 & 15.3 & 1.2 & 4.1 & 5.2 & 2.4 & 2.4 & 102 \\
4 & 200 & 0.3 & 12.6 & 16.9 & 13.6 & 18.1 & 20.2 & 1.9 & 8.1 & 9.3 & 5.2 & 5.2 & 78 \\
4 & 1000 & 0.1 & 5.3 & 6.4 & 5.9 & 7.1 & 54.1 & 4.0 & 3.8 & 4.6 & 3.8 & 3.4 & 243 \\
4 & 1000 & 0.3 & 6.0 & 7.7 & 6.8 & 8.8 & 58.3 & 5.3 & 3.0 & 2.8 & 3.8 & 3.1 & 225 \\
\hline
6 & 200 & 0.1 & 17.7 & 22.9 & 17.9 & 24.0 & 18.4 & 1.8 & 11.0 & 12.1 & 3.0 & 3.1 & 210 \\
6 & 200 & 0.3 & 24.8 & 31.9 & 26.6 & 34.9 & 22.0 & 2.8 & 15.1 & 16.2 & 4.4 & 4.4 & 208 \\
6 & 1000 & 0.1 & 11.6 & 14.6 & 13.7 & 17.5 & 71.7 & 6.0 & 6.5 & 4.9 & 4.1 & 3.3 & 287 \\
6 & 1000 & 0.3 & 15.3 & 18.5 & 18.3 & 22.8 & 84.5 & 12.3 & 7.0 & 5.2 & 4.5 & 3.1 & 347 \\
\hline
8 & 200 & 0.1 & 20.0 & 24.4 & 21.0 & 26.5 & 20.6 & 1.3 & 15.0 & 10.8 & 2.8 & 2.6 & 219 \\
8 & 200 & 0.3 & 21.2 & 25.9 & 22.2 & 26.5 & 21.9 & 1.8 & 17.0 & 13.4 & 2.2 & 2.1 & 257 \\
8 & 1000 & 0.1 & 22.9 & 26.7 & 24.4 & 29.4 & 92.6 & 15.1 & 14.2 & 10.5 & 5.1 & 3.5 & 312 \\
8 & 1000 & 0.3 & 32.6 & 40.7 & 35.4 & 45.2 & 140.8 & 30.9 & 19.3 & 13.1 & 7.8 & 5.7 & 415 \\
\hline
\end{tabular}
\begin{flushleft}
\footnotesize
\textit{Note.} Result for TS-MDFA is based on 10 multiple starts. Results for Bayes are based on 11,000 sampling iterations for each of two independent Markov chains. $M$ = mean; $SD$ = standard deviation.
\end{flushleft}
}
\end{table}

\begin{table}[htbp]
\textcolor{black}{
\centering
\caption{Required computation time (in seconds) across estimation methods ($\beta = 0.5$; restricted to cases in which no improper solutions occurred for each method).}
\label{tab:computation_time_proper_only_beta_0p5}
\begin{tabular}{ccc|rrrrrrrrrr|r}
\hline
\multicolumn{3}{c|}{} &
\multicolumn{2}{c}{ML} &
\multicolumn{2}{c}{CML} &
\multicolumn{2}{c}{Bayes} &
\multicolumn{2}{c}{ULS} &
\multicolumn{2}{c|}{TS-MDFA} &
$n$ \\
\hline
$T$ & $N$ & $\psi^2$ &
$M$ & $SD$ &
$M$ & $SD$ &
$M$ & $SD$ &
$M$ & $SD$ &
$M$ & $SD$ &
\\
\hline
4 & 200 & 0.1 & 8.4 & 8.9 & 10.0 & 14.3 & 19.6 & 1.0 & 9.0 & 13.2 & 4.6 & 4.4 & 132 \\
4 & 200 & 0.3 & 9.7 & 12.4 & 14.2 & 17.8 & 20.4 & 1.7 & 8.4 & 9.8 & 3.7 & 4.4 & 82 \\
4 & 1000 & 0.1 & 5.1 & 6.2 & 6.5 & 9.3 & 50.8 & 2.8 & 3.5 & 4.5 & 3.1 & 2.6 & 341 \\
4 & 1000 & 0.3 & 6.5 & 8.8 & 7.5 & 10.5 & 57.3 & 5.5 & 3.4 & 4.2 & 3.8 & 3.2 & 302 \\
\hline
6 & 200 & 0.1 & 18.8 & 24.1 & 22.1 & 27.8 & 21.2 & 1.1 & 17.7 & 19.9 & 4.5 & 4.8 & 288 \\
6 & 200 & 0.3 & 20.2 & 24.1 & 22.8 & 28.4 & 22.5 & 1.7 & 16.2 & 16.7 & 4.5 & 4.2 & 324 \\
6 & 1000 & 0.1 & 11.2 & 14.4 & 13.9 & 17.8 & 60.3 & 3.3 & 6.8 & 4.9 & 3.1 & 2.6 & 428 \\
6 & 1000 & 0.3 & 13.5 & 17.6 & 15.1 & 19.2 & 68.9 & 9.4 & 6.8 & 5.9 & 3.8 & 2.8 & 476 \\
\hline
8 & 200 & 0.1 & 31.1 & 40.5 & 38.0 & 50.6 & 22.4 & 1.8 & 25.7 & 19.6 & 3.6 & 3.5 & 335 \\
8 & 200 & 0.3 & 31.8 & 41.0 & 37.3 & 46.2 & 25.1 & 1.8 & 26.5 & 20.6 & 4.3 & 3.7 & 401 \\
8 & 1000 & 0.1 & 29.3 & 36.1 & 32.0 & 38.8 & 93.6 & 11.8 & 21.5 & 16.6 & 5.1 & 3.6 & 451 \\
8 & 1000 & 0.3 & 20.2 & 25.7 & 21.8 & 27.3 & 75.5 & 10.7 & 13.4 & 9.0 & 3.4 & 1.8 & 480 \\
\hline
\end{tabular}
\begin{flushleft}
\footnotesize
\textit{Note.} Result for TS-MDFA is based on 10 multiple starts. Results for Bayes are based on 11,000 sampling iterations for each of two independent Markov chains. $M$ = mean; $SD$ = standard deviation.
\end{flushleft}
}
\end{table}

\begin{table}[tb]
\textcolor{black}{
  \centering
  \caption{Point estimates and standard errors for parameters in the STARTS model obtained from each estimation method ($\sigma_{w1}^2$ included as a free parameter). Standard errors in the Bayesian estimation correspond to posterior standard deviations. The values shown in bold indicate improper solutions.}
  \label{tab:starts_estimates_five_parameters}
  \begin{tabular}{c*{4}{cc}cc}
    \hline
    & \multicolumn{2}{c}{ML} 
    & \multicolumn{2}{c}{CML}
    & \multicolumn{2}{c}{ULS}
    & \multicolumn{2}{c}{Bayes}
    & \multicolumn{2}{c}{TS-MDFA} \\
    \cline{2-11}
    Parameter 
      & Est. & SE 
      & Est. & SE 
      & Est. & SE 
      & Est. & SD 
      & Est. & SE\\
    \hline
    $\phi^2$ 
      & 0.114 & 0.015 
      & 0.091 & 0.017 
      & 0.015 & 0.034
      & 0.089 & 0.018
      & 0.089 & 0.022 \\
    $\psi^2$ 
      & \bd{-0.304} & 0.124 
      & \bd{0.000}  & NA    
      & 0.134 & 0.055 
      & 0.013 & 0.012    
      & 0.030 & 0.034 \\
    $\sigma_{w1}^2$ 
      & 0.582 & 0.125 
      & 0.300 & 0.020 
      & 0.270 & 0.040 
      & 0.292 & 0.024     
      & 0.241 & 0.042 \\
    $\omega^2$ 
      & 0.845 & 0.129 
      & 0.518 & 0.013 
      & 0.359 & 0.050 
      & 0.504 & 0.019   
      & 0.478 & 0.047 \\
    $\beta$ 
      & 0.251 & 0.048 
      & 0.442 & 0.024 
      & 0.648 & 0.071
      & 0.456 & 0.029 
      & 0.476 & 0.035 \\
    \hline
  \end{tabular}
}
\end{table}

\newpage
\thispagestyle{empty}
\renewcommand{\thefigure}{S\arabic{figure}}
\begin{figure}[htbp]
\textcolor{black}{
\centering
\includegraphics[width=\linewidth]{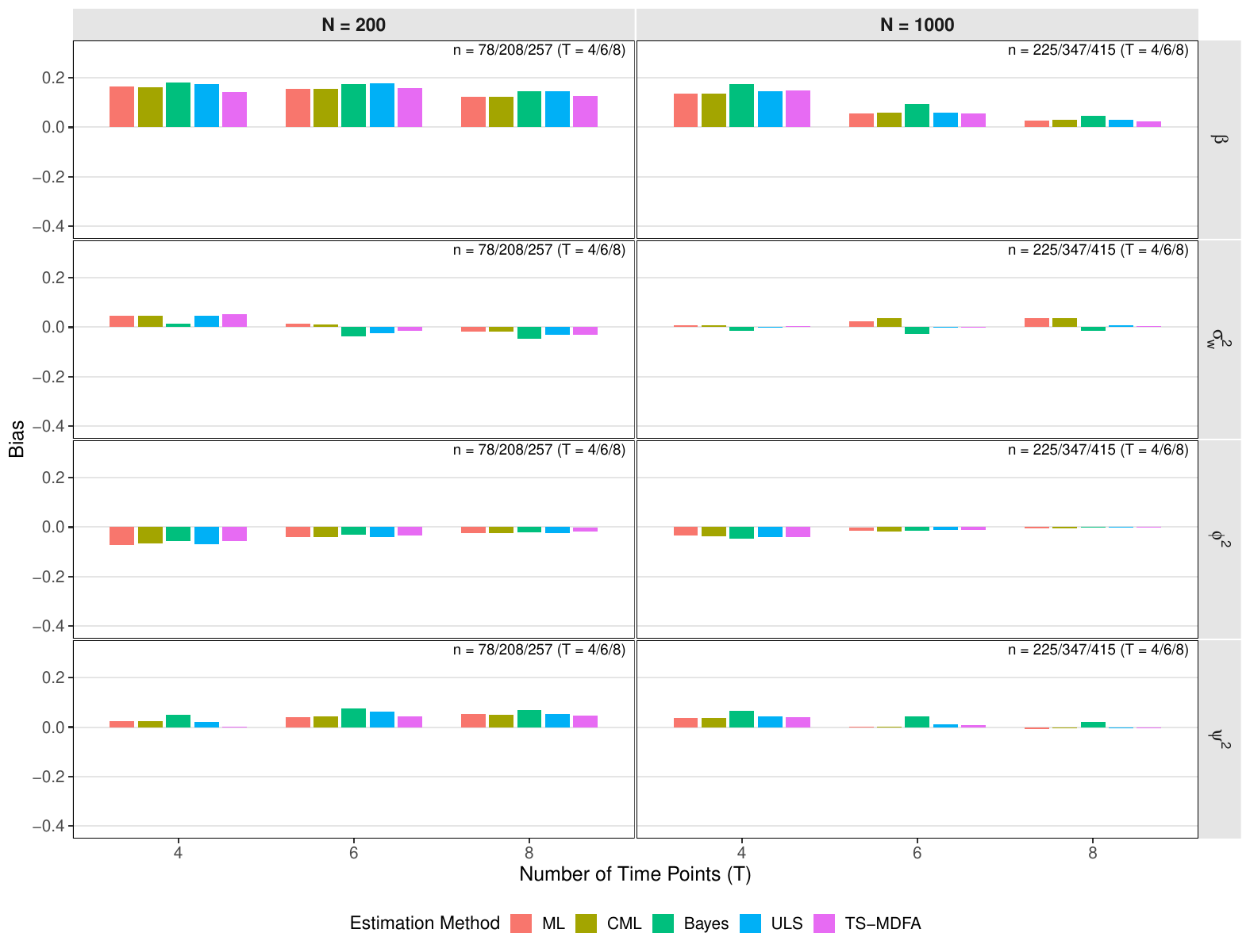}
\caption{Bias estimates by parameter type across methods ($\beta=0.3$, $\psi^2 = 0.3$). \textit{Note.} The number shown in the upper-right corner of each panel indicates the number of cases used to compute bias, i.e., the number of cases in which admissible solutions were observed for all methods.}
}
\end{figure}

\thispagestyle{empty}
\renewcommand{\thefigure}{S\arabic{figure}}
\begin{figure}[htbp]
\textcolor{black}{
\centering
\includegraphics[width=\linewidth]{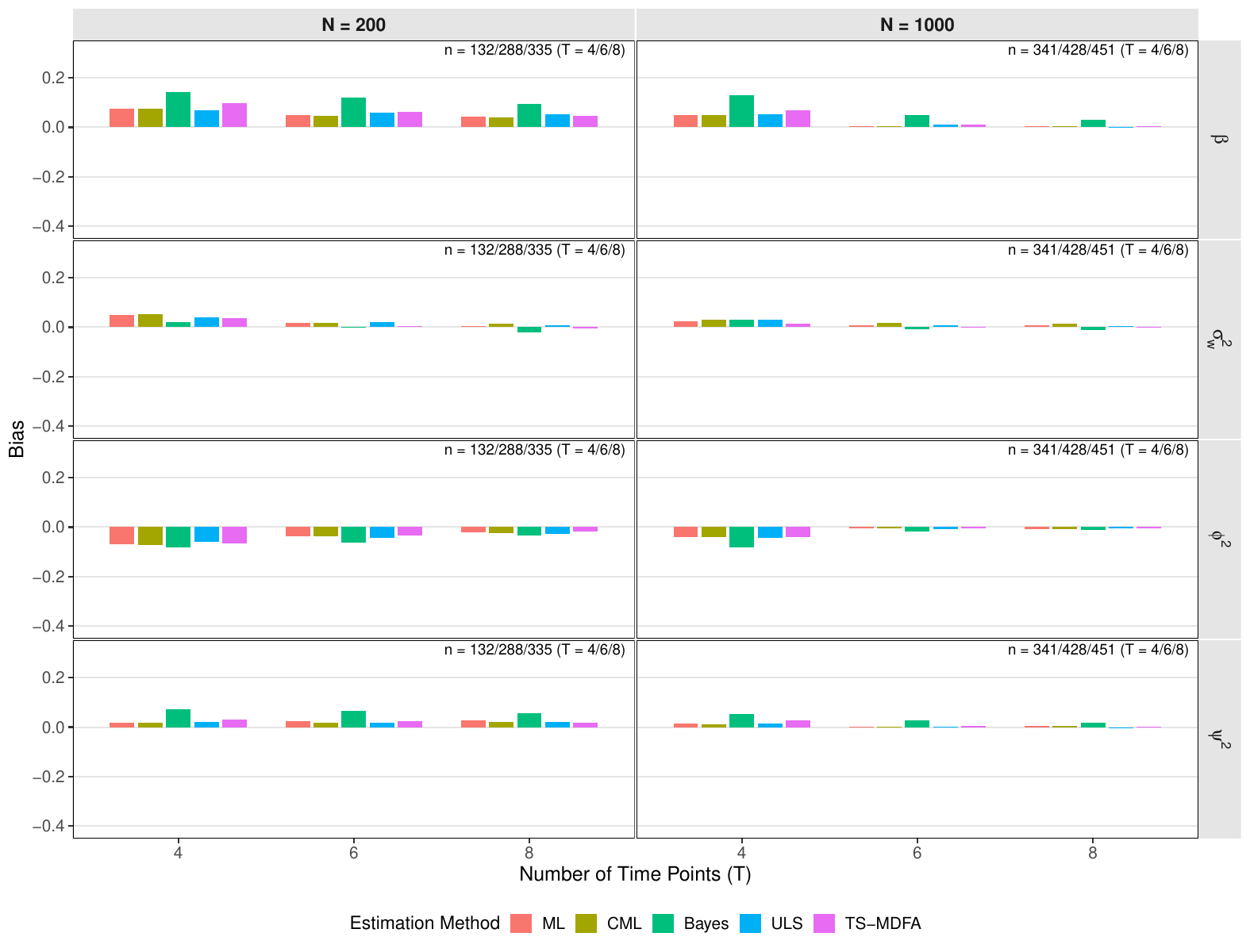}
\caption{Bias estimates by parameter type across methods ($\beta=0.5$, $\psi^2 = 0.1$). \textit{Note.} The number shown in the upper-right corner of each panel indicates the number of cases used to compute bias, i.e., the number of cases in which admissible solutions were observed for all methods.}
}
\end{figure}

\thispagestyle{empty}
\renewcommand{\thefigure}{S\arabic{figure}}
\begin{figure}[htbp]
\textcolor{black}{
\centering
\includegraphics[width=\linewidth]{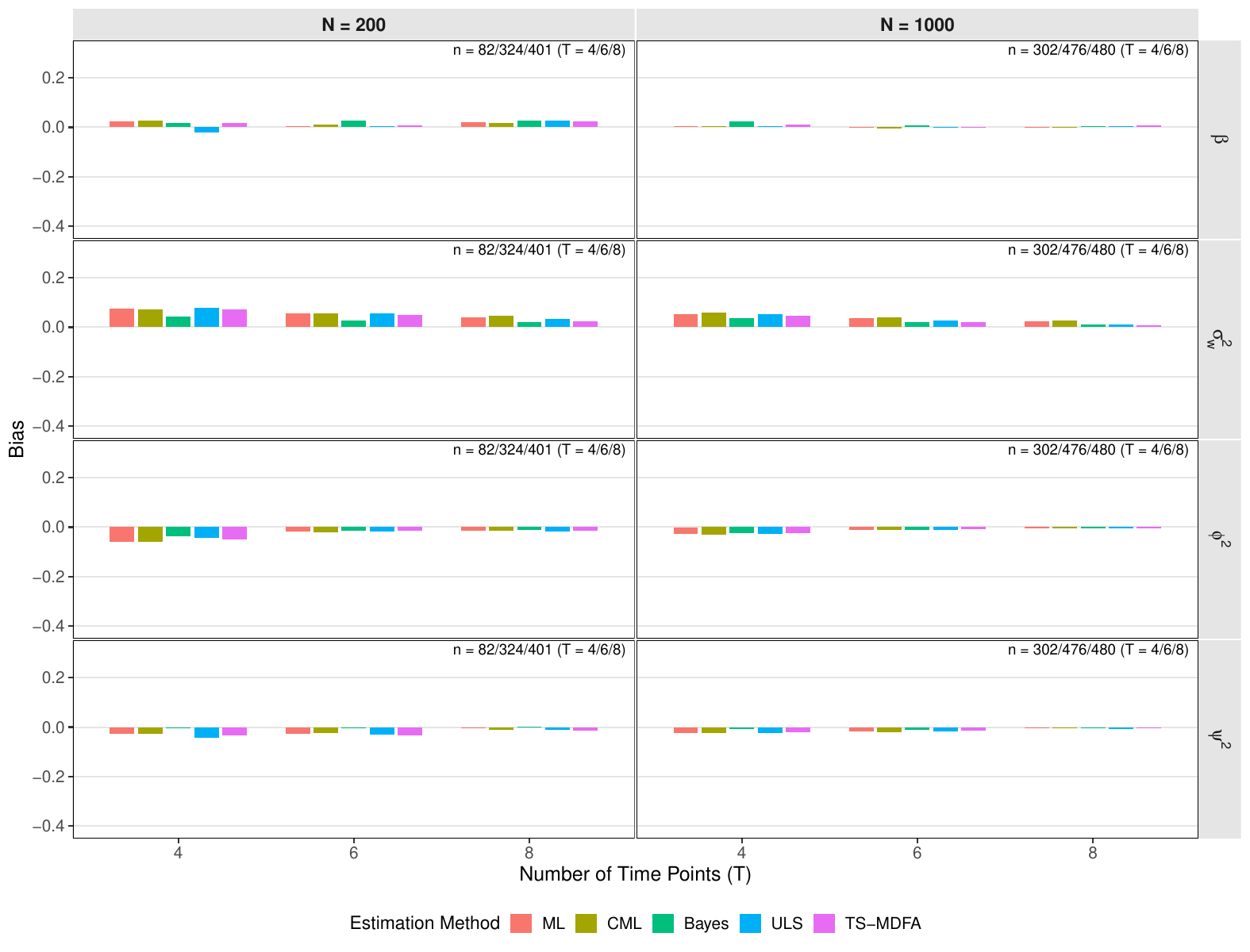}
\caption{Bias estimates by parameter type across methods ($\beta=0.5$, $\psi^2 = 0.3$). \textit{Note.} The number shown in the upper-right corner of each panel indicates the number of cases used to compute bias, i.e., the number of cases in which admissible solutions were observed for all methods.}
}
\end{figure}

\thispagestyle{empty}
\renewcommand{\thefigure}{S\arabic{figure}}
\begin{figure}[htbp]
\textcolor{black}{
\centering
\includegraphics[width=\linewidth]{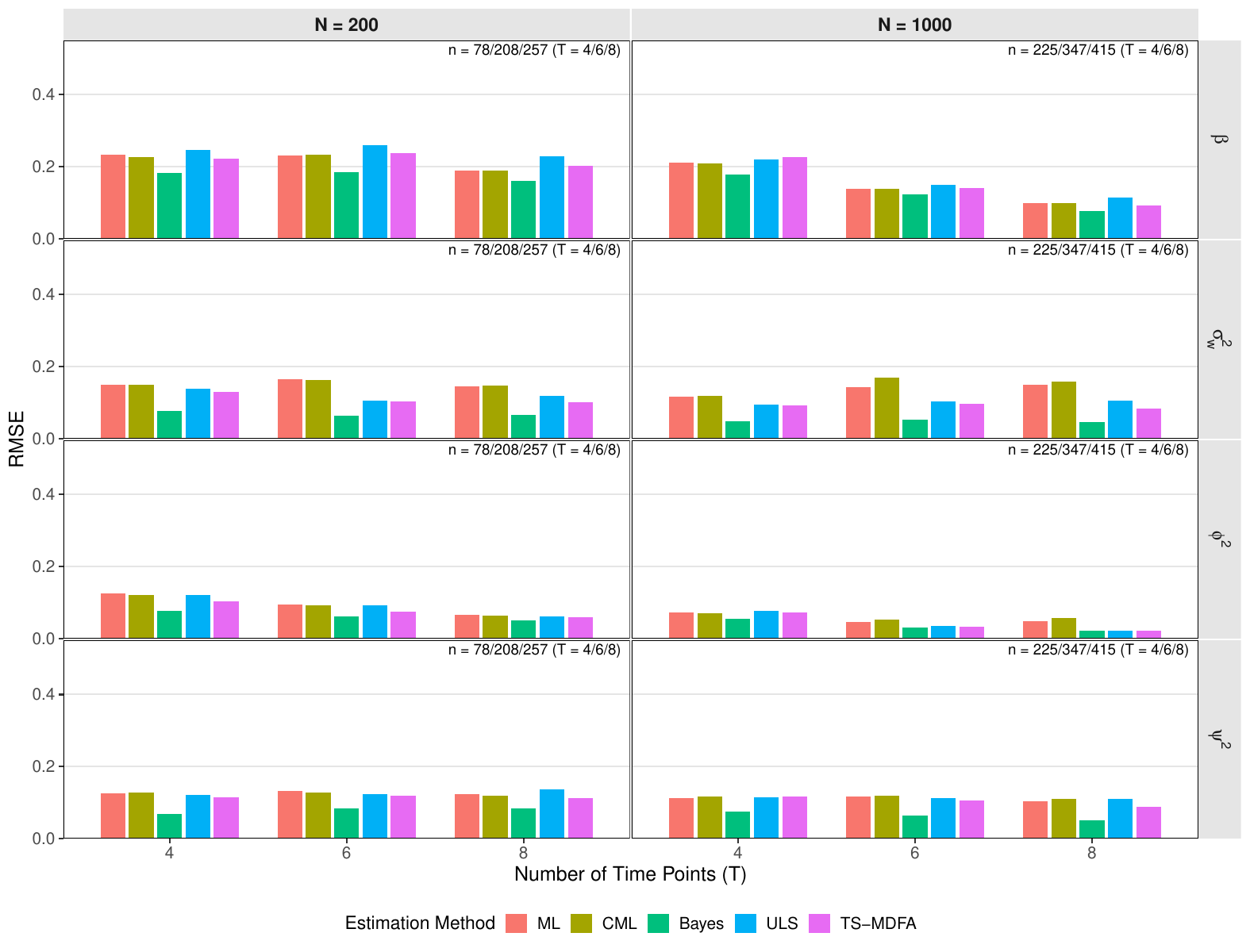}
\caption{RMSE estimates by parameter type across methods ($\beta=0.3$, $\psi^2 = 0.3$). \textit{Note.} The number shown in the upper-right corner of each panel indicates the number of cases used to compute RMSE, i.e., the number of cases in which admissible solutions were observed for all methods.}
}
\end{figure}

\thispagestyle{empty}
\renewcommand{\thefigure}{S\arabic{figure}}
\begin{figure}[htbp]
\textcolor{black}{
\centering
\includegraphics[width=\linewidth]{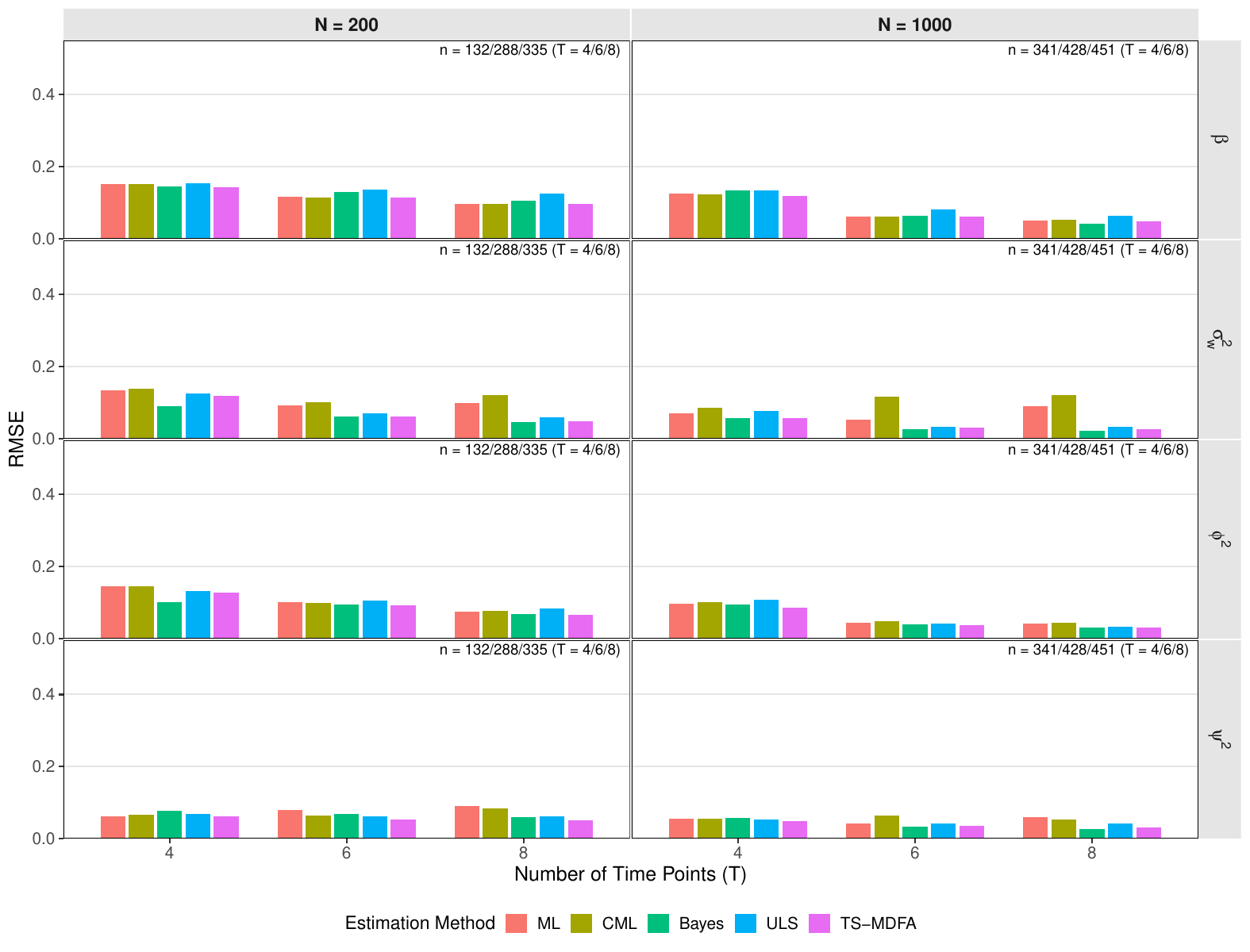}
\caption{RMSE estimates by parameter type across methods ($\beta=0.5$, $\psi^2 = 0.1$). \textit{Note.} The number shown in the upper-right corner of each panel indicates the number of cases used to compute RMSE, i.e., the number of cases in which admissible solutions were observed for all methods.}
}
\end{figure}

\thispagestyle{empty}
\renewcommand{\thefigure}{S\arabic{figure}}
\begin{figure}[htbp]
\textcolor{black}{
\centering
\includegraphics[width=\linewidth]{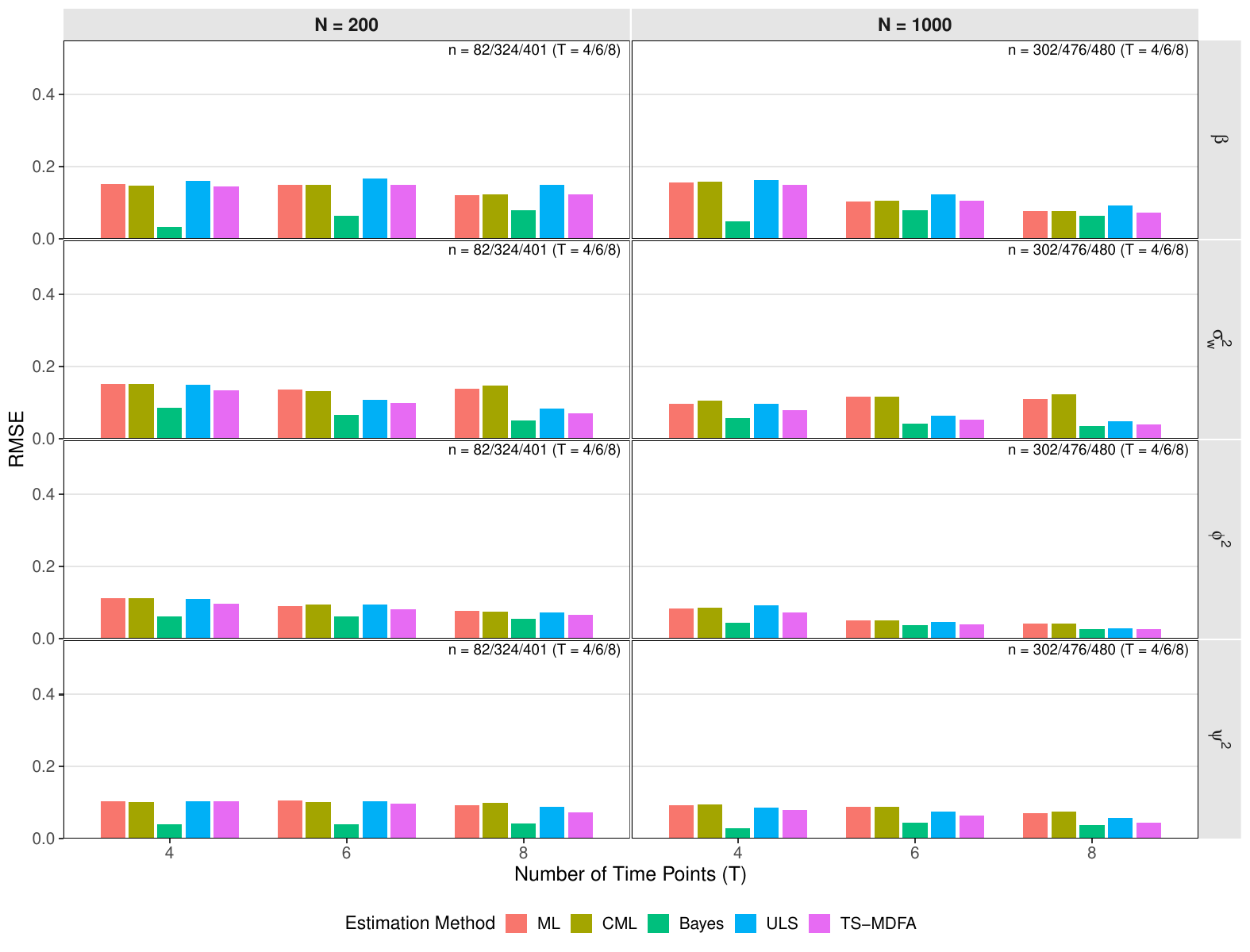}
\caption{RMSE estimates by parameter type across methods ($\beta=0.5$, $\psi^2 = 0.3$). \textit{Note.} The number shown in the upper-right corner of each panel indicates the number of cases used to compute RMSE, i.e., the number of cases in which admissible solutions were observed for all methods.}
}
\end{figure}

\thispagestyle{empty}
\renewcommand{\thefigure}{S\arabic{figure}}
\begin{figure}[htbp]
\textcolor{black}{
\centering
\includegraphics[width=\linewidth]{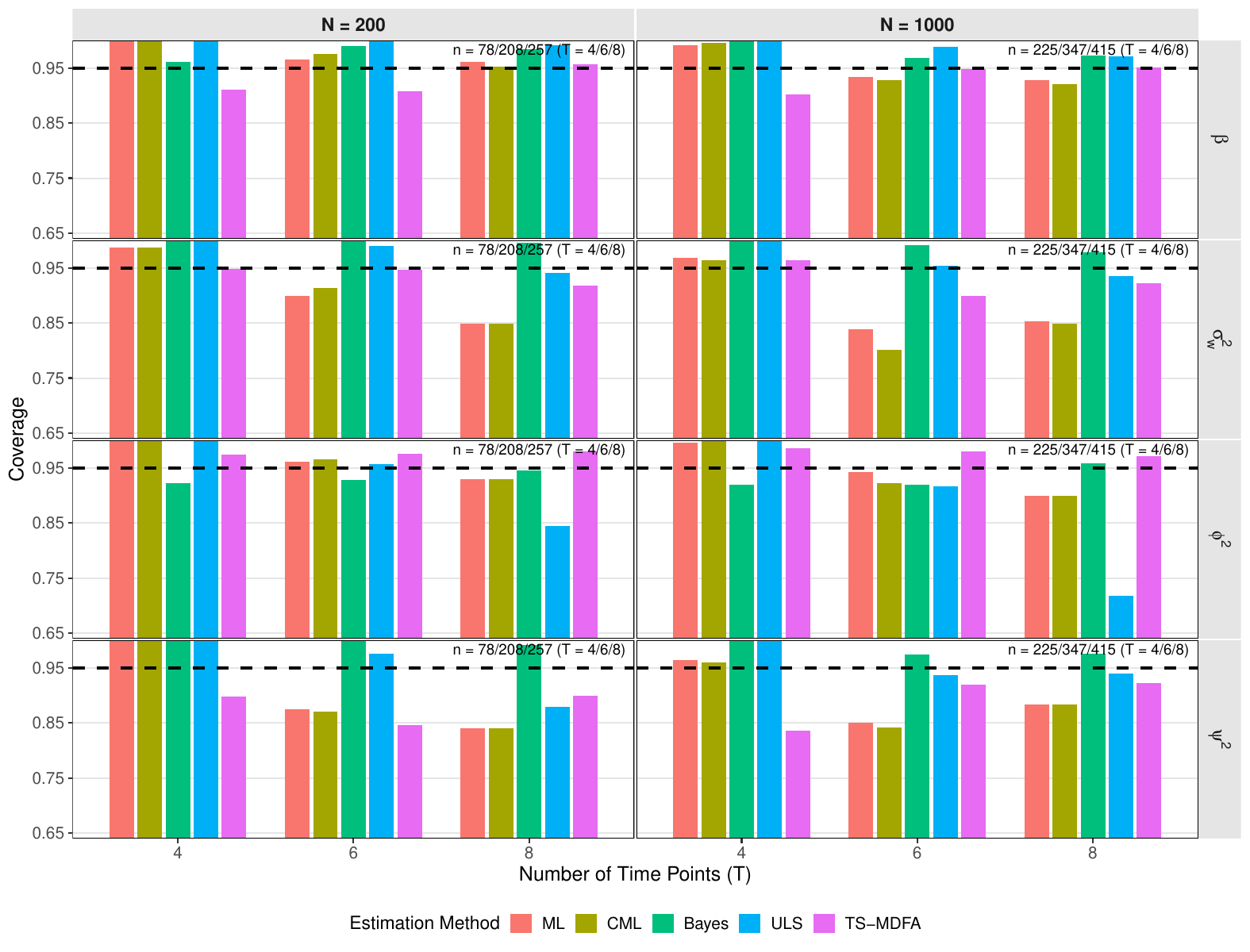}
\caption{Coverages by parameter type across methods ($\beta=0.3$, $\psi^2 = 0.3$). \textit{Note.} The number shown in the upper-right corner of each panel indicates the number of cases used to compute coverages, i.e., the number of cases in which admissible solutions were observed for all methods.}
}
\end{figure}

\thispagestyle{empty}
\renewcommand{\thefigure}{S\arabic{figure}}
\begin{figure}[htbp]
\textcolor{black}{
\centering
\includegraphics[width=\linewidth]{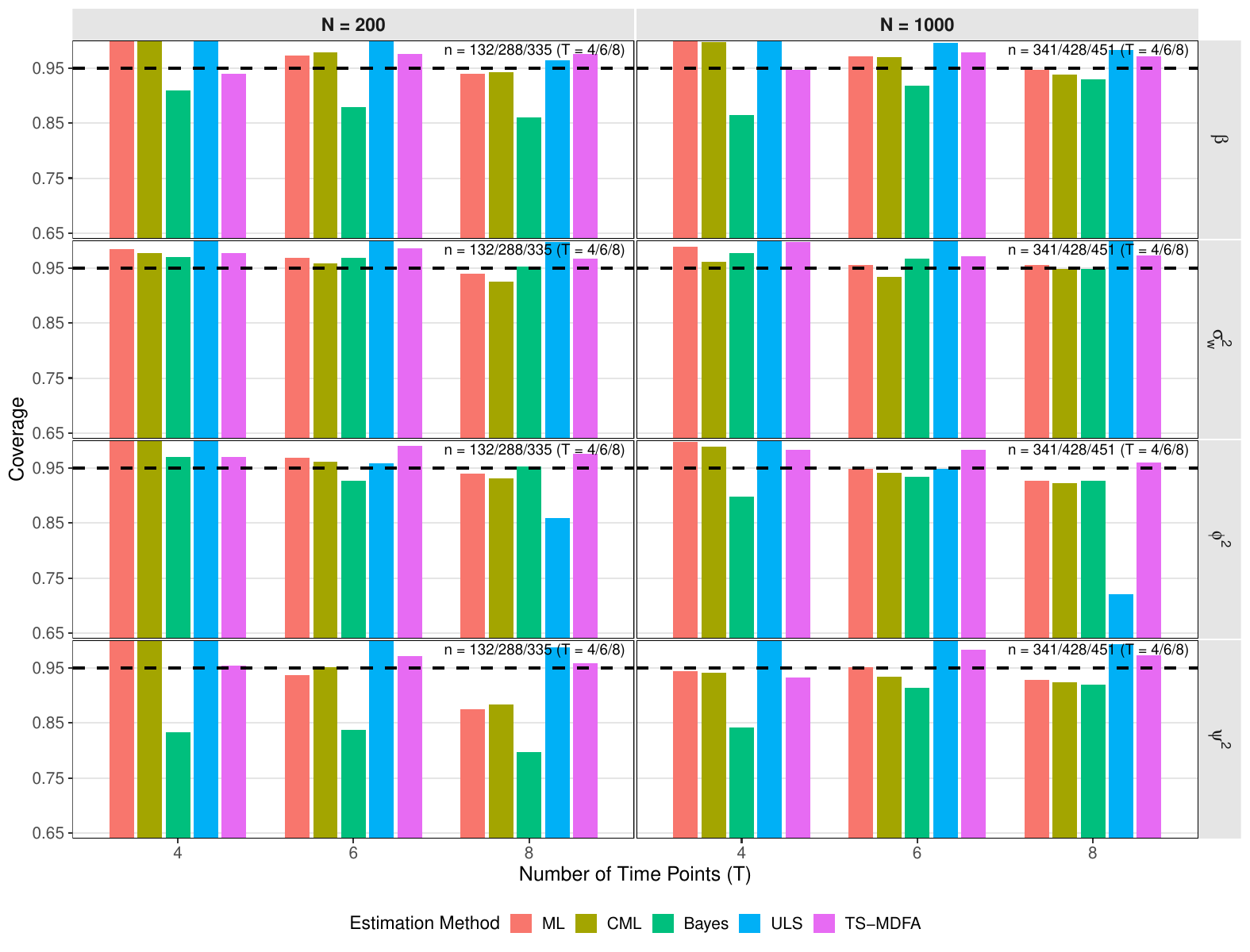}
\caption{Coverages by parameter type across methods ($\beta=0.5$, $\psi^2 = 0.1$). \textit{Note.} The number shown in the upper-right corner of each panel indicates the number of cases used to compute coverages, i.e., the number of cases in which admissible solutions were observed for all methods.}
}
\end{figure}

\thispagestyle{empty}
\renewcommand{\thefigure}{S\arabic{figure}}
\begin{figure}[htbp]
\textcolor{black}{
\centering
\includegraphics[width=\linewidth]{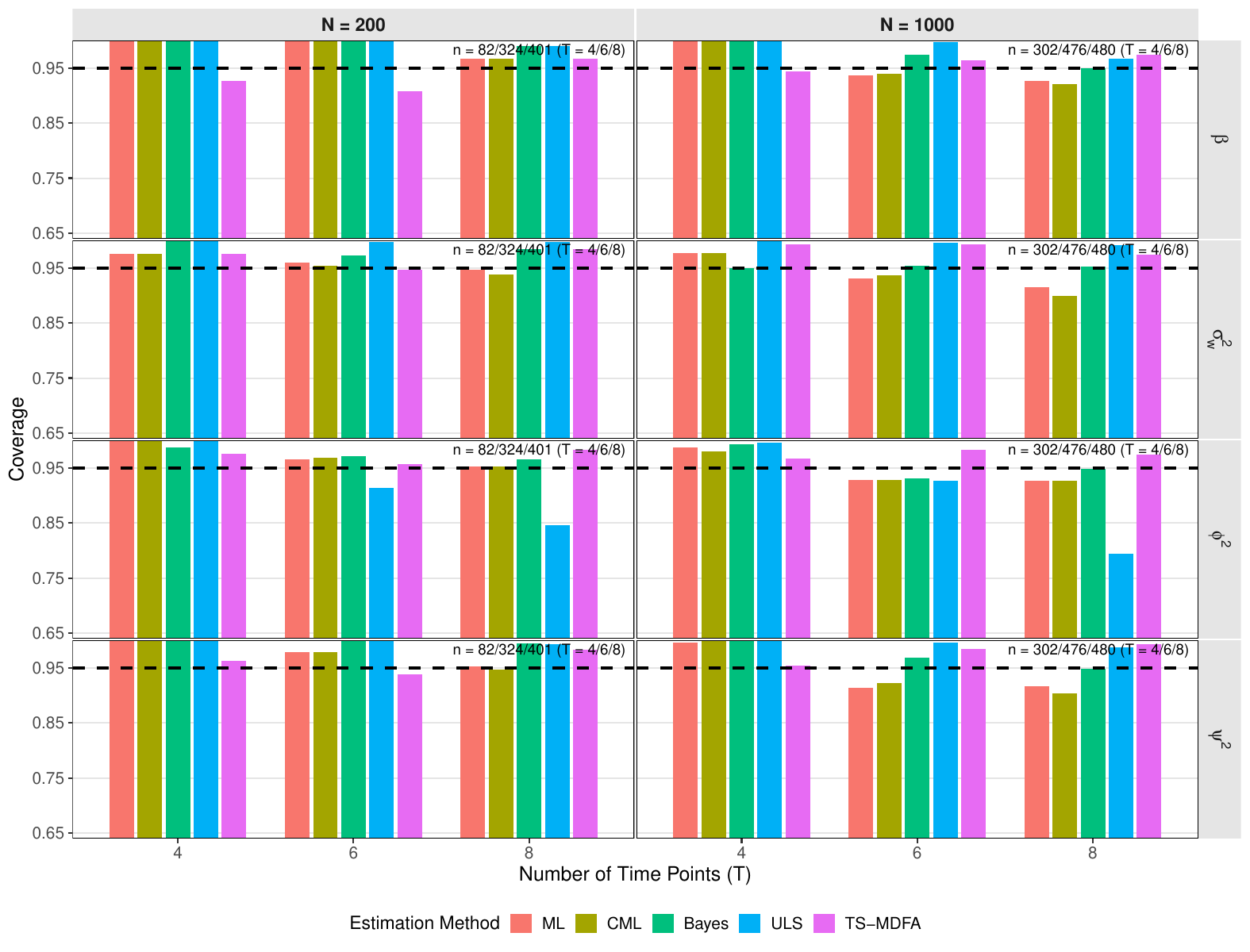}
\caption{Coverages by parameter type across methods ($\beta=0.5$, $\psi^2 = 0.3$). \textit{Note.} The number shown in the upper-right corner of each panel indicates the number of cases used to compute coverages, i.e., the number of cases in which admissible solutions were observed for all methods.}
}
\end{figure}

\thispagestyle{empty}
\renewcommand{\thefigure}{S\arabic{figure}}
\begin{figure}[htbp]
\textcolor{black}{
\centering
\includegraphics[width=\linewidth]{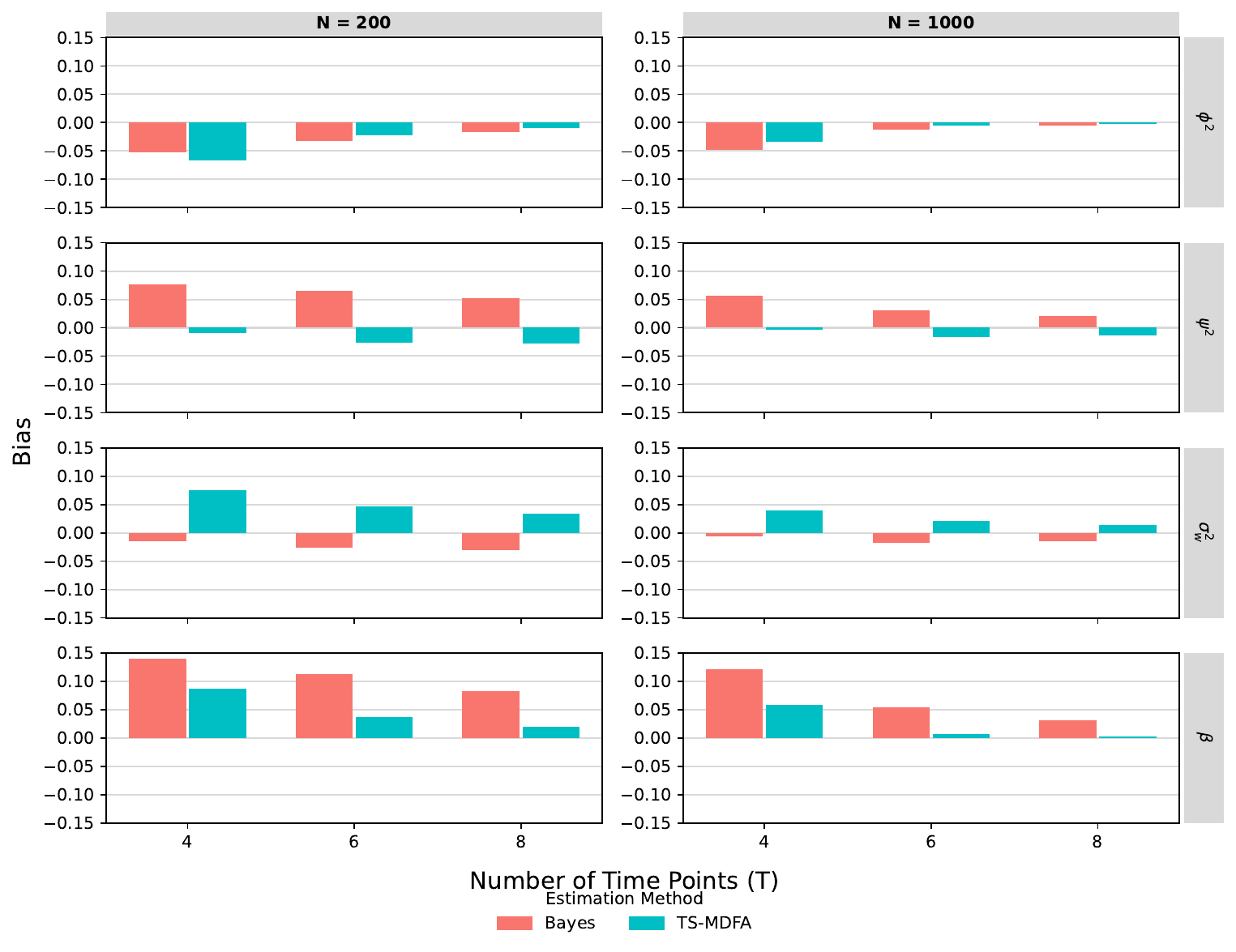}
\caption{Bias estimates by parameter type in Bayesian estimation and TS-MDFA (computed using all 500 replications).}
}
\end{figure}

\newpage
\thispagestyle{empty}
\renewcommand{\thefigure}{S\arabic{figure}}
\begin{figure}[htbp]
\textcolor{black}{
\centering
\includegraphics[width=\linewidth]{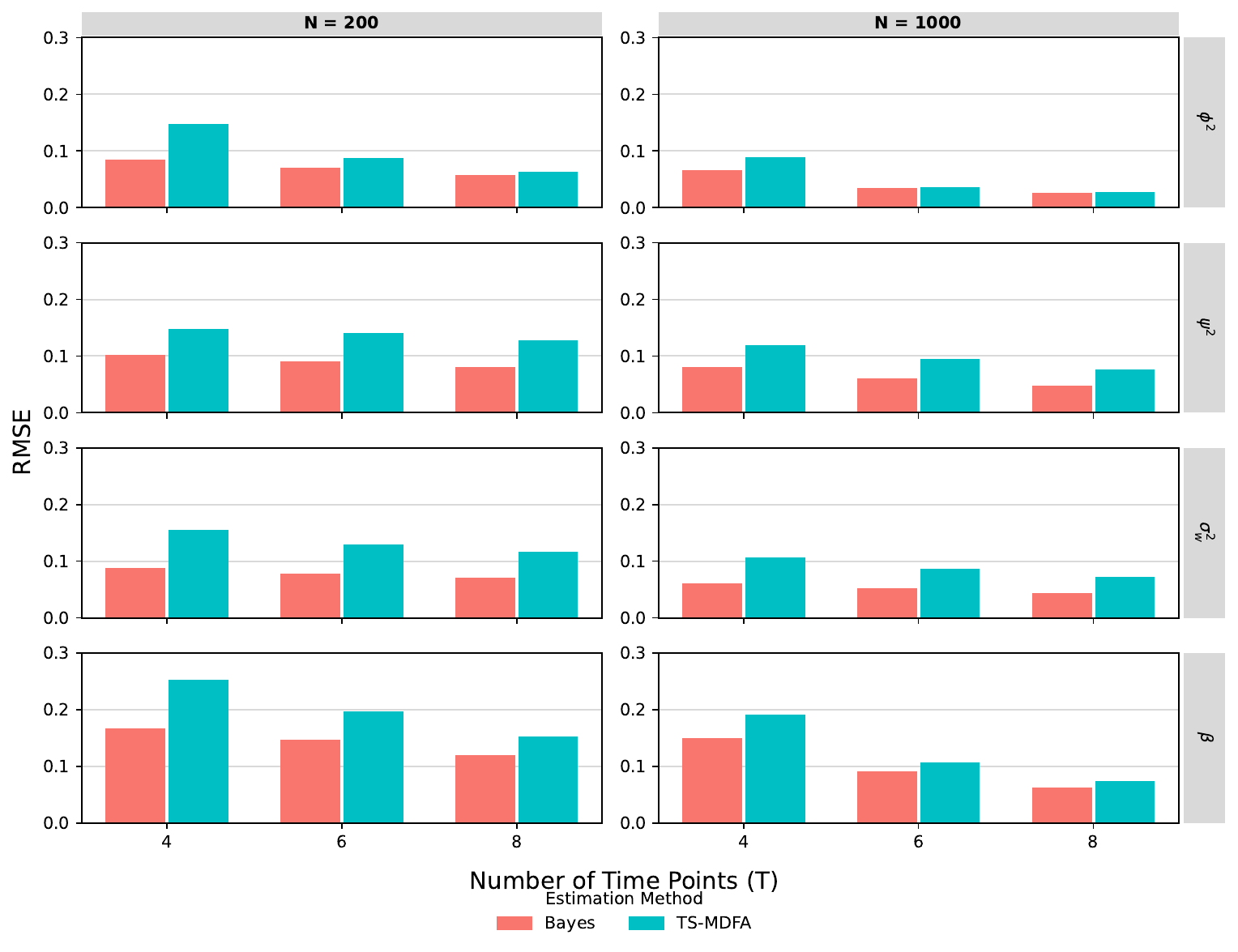}
\caption{RMSE estimates by parameter type in Bayesian estimation and TS-MDFA (computed using all 500 replications).}
}
\end{figure}

\newpage
\thispagestyle{empty}
\renewcommand{\thefigure}{S\arabic{figure}}
\begin{figure}[htbp]
\textcolor{black}{
\centering
\includegraphics[width=\linewidth]{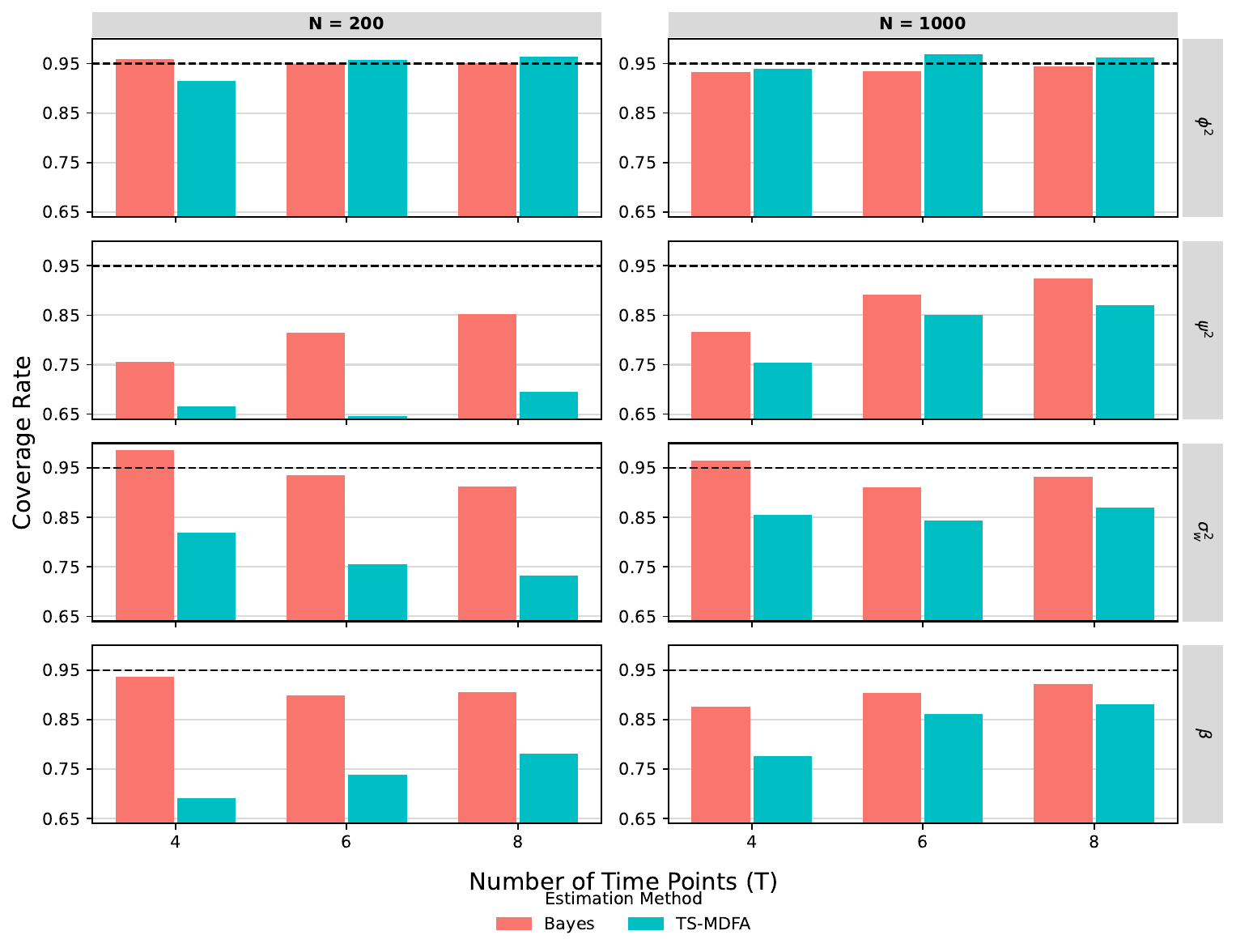}
\caption{Coverages by parameter type in Bayesian estimation and TS-MDFA (computed using all 500 replications).}
}
\end{figure}

\newpage
\thispagestyle{empty}
\renewcommand{\thefigure}{S\arabic{figure}}
\begin{figure}[htbp]
\textcolor{black}{
\centering
\includegraphics[width=\linewidth]{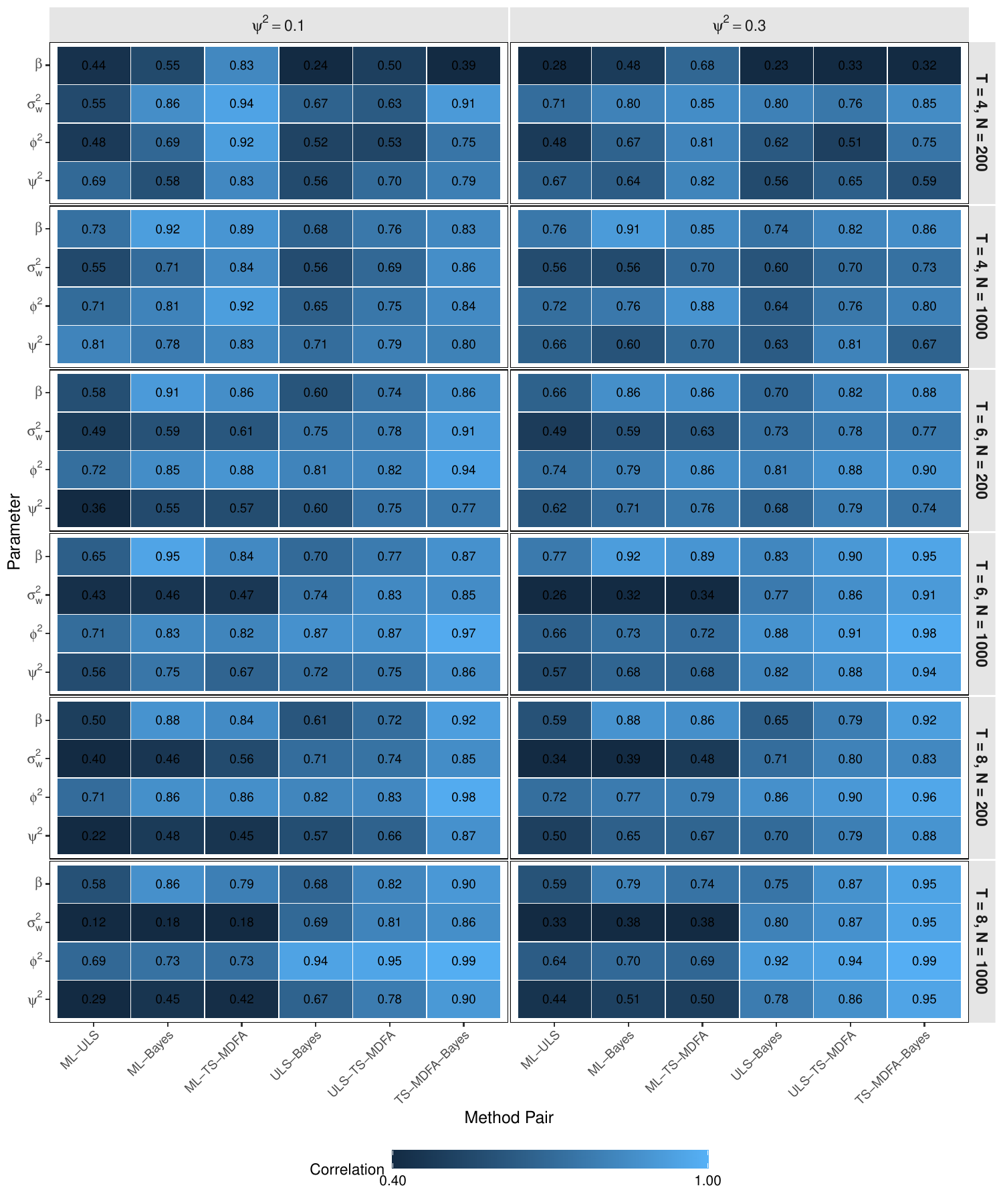}
\caption{Correlations of parameter estimates across method pairs ($\beta=0.5$).}
}
\end{figure}

\end{document}